\documentclass[twocolumn,prc,showpacs,preprintnumbers,
               unsortedaddress,amsmath,amssymb,floatfix]{revtex4}
\usepackage{graphicx}
\usepackage{dcolumn}
\usepackage{bm}
\usepackage{longtable}

\newcommand{\beq}{\begin{equation}}
\newcommand{\eeq}{\end{equation}}
\newcommand{\beqn}{\begin{eqnarray}}
\newcommand{\eeqn}{\end{eqnarray}}
\newcommand{\btab}{\begin{tabular}}
\newcommand{\etab}{\end{tabular}}

\newcommand{\etal}{{\em{et al.}}}

\newcommand{\ls}{\left[}
\newcommand{\rs}{\right]}

\usepackage{multirow,amsmath,booktabs}

\begin{document}

\title{Equation of State of nuclear matter in a Virial expansion of nucleons and nuclei}

\author{G.~Shen\footnote{e-mail:  gshen@indiana.edu}}
\affiliation{Nuclear Theory Center and Department of Physics,
Indiana University Bloomington, IN 47405}
\author{C.~J.~Horowitz\footnote{e-mail:
horowit@indiana.edu} } \affiliation{Nuclear Theory Center and
Department of Physics, Indiana University Bloomington, IN 47405}
\author{S.~Teige\footnote{e-mail:
steige@indiana.edu}}
\affiliation{University Information Technology Services, Indiana University, Bloomington, IN 47408}
\date{\today}

\begin{abstract}
We study the equation of state (EOS) of nuclear
matter at subnuclear density in a Virial expansion for a nonideal gas. The gas consists of neutrons, protons, alpha particles, and 8980 species of nuclei with $A \ge 12$ and masses from the finite-range droplet model (FRDM).  At very low density, the Virial
expansion reduces to nuclear statistical equilibrium.  At higher density, the Virial results match smoothly to the relativistic mean field results discussed in our previous paper.   We tabulate the resulting EOS at over 73,000 grid points in the temperature
range $T$ = 0.158 to 15.8 MeV, the density range $n_B$ = 10$^{-8}$
to 0.1 fm$^{-3}$, and the proton fraction range $Y_P$ = 0.05 to 0.56.  In the future we plan to match these low density results to our earlier high density mean field results, and generate a full EOS table for use in supernova and neutron star merger simulations.  This Virial EOS is exact in the low density limit.
\end{abstract}


\maketitle

\section{Introduction}

One of the main ingredients in simulations \cite{simulation1,simulation2} of core collpase supernovae and neutron star mergers is the Equation of State (EOS) for hot dense nuclear matter.  The EOS, along with detailed information on the composition of nuclear matter, may play an important role in the neutrino-matter dynamics \cite{Burrows04} and supernova explosion mechanism.   In a previous paper \cite{SHT10a} we used a relativistic mean field (RMF) model to calculate the free-energy of non-uniform matter at intermediate density and uniform matter at high density.   

Simulations of the core collapse and explosion of supernovae depend heavily on the EOS at subnuclear density, especially the detailed composition.  In this work, we study subnuclear density nuclear matter in a Virial expansion for a nonideal gas, consisting of neutrons, protons, alpha particles, and 8980 species of heavy nuclei ($A\geq 12$) with masses from the finite-range droplet model (FRDM) \cite{FRDM}.  The Virial results will cover  the density range $n_B$ = 10$^{-8}$ to 0.1 fm$^{-3}$, the temperature
range $T$ = 0.158 to 15.8 MeV, and the proton fraction range $Y_P$ = 0.05
to 0.56.  (For temperature higher than 15.8 MeV, matter is uniform and fully described in the RMF model \cite{SHT10a}.) The distribution of nuclei for given conditions is obtained in this approach, while the existing EOS tables of Lattimer-Swesty (L-S) \cite{LS} and H. Shen, Toki, Oyamatsu and Sumiyoshi (S-S) \cite{Shen98a,Shen98}, use a single heavy nucleus approximation.  Our Virial EOS will be matched, using a thermodynamically consistent interpolation scheme, to the RMF EOS obtained in our previous paper \cite{SHT10a} to generate a full EOS table for supernova.

Detailed information on the distributions of nuclei in the EOS table is important for neutrino-matter dynamics. Neutrinos radiate 99\% of the energy released in supernovae. Besides gravitational wave signals, neutrinos are the only messenger through which one can directly probe the EOS inside supernovae.  The neutrinosphere is the surface of last scattering before  $\nu$ or $\bar{\nu}$ escape.  The neutrinosphere is expected to occur at a density around\ $10^{11}$ g/cm$^3$ and this is consistent with the available information from a handful
events in SN1987a  \cite{SN1987a}.  The composition of matter at subnuclear density constrains the position of the neutrinosphere and influences the spectra of emitted neutrinos and antineutrinos. For example, in a recent study \cite{light07}, light nuclei with mass 2, 3 and 4 were found to have an important influence on the spectra of anti-electron neutrinos.


For matter at subnuclear density and low entropy, heavy nuclei
tend to form. Nuclear statistical equilibrium (NSE) models treat
low density nuclear matter as a system of noninteracting nuclei in
statistical equilibrium, taking into account the binding
properties of heavy nuclei.  This has been widely used in nuclear
astrophysics \cite{NSE}. Recently, there have been several NSE based studies
of the supernova EOS, see for example Refs.
\cite{NSE09} and \cite{Hempel09}.  These studies use modern mass tables with
thousands of nuclei and include excluded volume effects \cite{Hempel09}. The NSE models have the advantage of generating thermodynamically consistent EOS tables. However they also have several disadvantages.  First, NSE models themselves  can not be used to describe well the non-uniform matter at nearly nuclear density, when exotic pasta phases may appear. To generate a complete EOS table, NSE models need to be matched to uniform matter models at high density.  Thus low and high density matter are usually described with different models. Second, NSE models do not provide a systematic way to include the strong
interactions between nuclei.  These interactions could be important as
the density increases. Moreover, neutron matter at low density is closely
analogous to a unitary gas, whose properties can not be described
satisfactorily in either NSE models or normal mean field approaches (see for example
\cite{unitarygas}).  The Virial expansion, on the other hand, has been successfully used to describe the properties of unitary gases \cite{Ho04} as well as neutron
matter at low density \cite{HS05b}.


In a previous work \cite{HS05} on low density nuclear matter consisting of neutrons, protons and alpha ($\alpha$) particles, the Virial expansion has been used to
systematically incorporate second virial corrections from
nucleon-nucleon (N-N), N$\alpha$ and $\alpha\alpha$ elastic
scattering.  These light nuclei components are expected to
dominate in matter with high entropy.  The effects of second virial
corrections are found to be important for the composition and for the equation of state. In this work we use the  Virial expansion for a nonideal gas consisting of neutrons, protons, alpha particles, and thousands of heavy nuclei. The heavy nuclei included in this work are from FRDM mass tables \cite{FRDM}, which have 8979 nuclei with $A \geq$ 16. We also include $^{12}$C in the mass table. As mentioned above, the mass 2 and 3 nuclei, and other $A<12$ nuclei asside from $^4$He may be important for the spectra of
anti-electron neutrinos.  The inclusion of these light elements will
require information on second virial corrections among them and
nucleons, which are under current investigation \cite{mass3}.  In
this work the light elements are represented by alpha particles,
which typically have the largest abundance among the light elements in
statistical equilibrium.   At very low density, the Virial expansion
agrees with nuclear statistical equilibrium models.  As the density
rises, we find that the Virial gas matches smoothly to previous mean field results.

In this work we include the second order virial corrections among nucleons and alphas as in Ref.~\cite{HS05}.  Partition functions for heavy nuclei are included using the recipes of Fowler \etal \cite{Fowler78}.  In addition, some calculations are presented using partition functions based on the recipe of Rauscher \etal \cite{Rauscher97}  for comparison.  Equivalently, the partition functions can be considered as the sum of successive high orders of the Virial expansion for heavy nuclei. There have been many studies on the level density and partition
functions of hot nuclei in astrophysical environments.  For large scale
astrophysical applications, it is necessary to find both reliable
and computationally practical methods for the level density. Most of
these studies \cite{Gilbert65,Fowler78,Mazurek79,Engelbrecht91,Rauscher97} followed the original  non-interacting Fermi gas model of Hans Bethe \cite{Bethe36}.
For astrophysical nuclear reactions with temperature below a few
times 10$^{10} K$ \cite{Rauscher97,Rauscher00}, this
phenomenological approach gave excellent agreement with more
sophisticated Monte Carlo shell model calculations \cite{mc}, as
well as combinatorial approaches \cite{Engelbrecht91,Paar97}.  This justifies the application of the Fermi-gas description at and above the neutron separation energy.  For temperatures higher than a few times 10$^{10} K$, there are big ambiguities in
the values of the partition functions.  However, as suggested by some
authors \cite{NSE09} and supported by our own calculations,
these uncertainties have only a small impact on the thermodynamics of dense matter.

The effects of Coulomb interactions in the plasma can be estimated
by the plasma parameter $\Gamma_p = (Ze)^2/akT$, where $Z$
is the atomic number of the nucleus, $T$ is the temperature and $a$ is the
spacing between nuclei.  For matter at low density, $\Gamma_p$ is
smaller than one and the effect of Coulomb corrections is
small.  However for matter at higher density and when the dominant
species carry large charges, $\Gamma_p$ can be much greater than one and
the effect of Coulomb interactions should be taken into account. The Coulomb correction to the plasma has been studied analytically up to high $\Gamma_p$ by the cluster expansion \cite{coulomb}. Generally the correction due to
electron-ion interactions will reduce the free energy of the plasma and
eventually crystalize the matter at high density.  For simplicity, in this work
the Coulomb interactions between nuclei and electrons are included via a Wigner-Seitz approximation with effective ion spheres for each species of nuclei, wherein local electrical neutrality is maintained. This Wigner-Seitz approximation for the Coulomb correction will be compared with a more rigorous cluster expansion method.

Based on the above Virial expansion, we generate an equation of state table which covers the  range of temperatures, densities and proton fractions shown in Tab.~\ref{tab:phasespace}.

\begin{table}[h]
\centering \caption{Range of temperatures, densities and proton
fractions in the EoS table.} \label{tab:phasespace}\btab{cccc}
\hline \hline
Parameter & minimum & maximum & number of grids \\
 \hline
T [MeV] & 10$^{-0.8}$ & 10$^{1.2}$ & 20 \\

log$_{10}$(n$_B$) [fm$^{-3}$] &-8.0 & -1.0 & 71 \\

Y$_P$  & 0.05  & 0.56 & 52 \\

\hline \etab
\end{table}

There are 7 points in temperature (0.16, 0.26, 0.40, 0.63, 0.71, 0.79 and 0.89 MeV) for temperature below 1 MeV.  For higher temperatures we use a spacing of 0.1 in $\log_{10}( T/ \ls {\rm{MeV}} \rs )$. This is a total of 20 different temperatures from $T$ = 0.16 to 15.8 MeV. We use a spacing of 0.1 in $\log_{10}(n_B/ \ls {\rm{fm}}^{-3}\rs )$, giving a total of 71 points in density for $n_B = 10^{-8} \sim$ 0.1 fm$^{-3}$. We use a linear step 0.01 in proton fraction, giving a total of 52 points in proton fraction for Y$_P$ = 0.05 to 0.56. There is a total of 73,840 data points in the Virial gas calculation. This took 1,000 CPU days in Indiana University's supercomputer clusters.

The paper is organized as follows: in section \ref{formalism}
our Virial expansion for a nonideal gas is explained in detail. In section \ref{numerics} we
present the recipes for the nuclear partition function used in the Virial expansion. Section \ref{result} discusses the effect of different mass tables on the EOS, and shows several examples for the free energy and the distribution of nuclei in the Virial EOS. Section \ref{summary} presents a summary of our results and gives an outlook for future work.

\section{\label{formalism}Formalism}

We now describe our Virial expansion formalism for a gas consisting of neutrons (n), protons (p), alpha particles and heavy nuclei.  The grand partition function ${\cal Q}$ for a gas at pressure $P$ and volume $V$ is expanded to second order in the neutron fugacity $z_n$, the proton fugacity $z_p$ and the alpha particle fugacity $z_\alpha$ as follows, 
\beqn\label{partition1}
\frac{\mathrm{log}{\cal Q}}{V} &=&\ \frac{P}{T} = \frac 2
{\lambda_n^3}[z_n+z_p+(z_p^2+z_n^2)b_n+2z_pz_nb_{pn}]\cr && +\frac
1
{\lambda_\alpha^3}[z_\alpha+z_\alpha^2b_\alpha+2z_\alpha(z_n+z_p)b_{\alpha
n}]\cr && +\sum_{\substack{i}} \frac 1 {\lambda_i^3}z_i\Omega_i,
\eeqn 
where $\Omega_i$ is the partition function for nuclei and $b_n,
b_{pn}, b_\alpha,$ and $b_{\alpha n}$ are the second Virial
coefficients as defined in Ref.~\cite{HS05}.  The sum on $i$ runs over
different heavy nuclei, for which we use the FRDM mass table
\cite{FRDM} for A$\geq$ 16 and include $^{12}$C.  The thermal wavelength for species $a$ is $\lambda_a$, 
\beq\label{thermallength}
\lambda_a\ =\ \sqrt{2\pi/m_aT},\ a\ =\ n,\ p,\ \alpha,\ {\rm nuclei}.
\eeq 
From now on $i, j, ...$ are used for sums over heavy nuclei and $a, b,
...$ are used for sums over all species.

There exist several recipes for the nuclear partition function
$\Omega_i$.  We will use that of Fowler \etal \cite{Fowler78} in
this work.  We also consider the choice of Rauscher \etal \cite{Rauscher97} as an option. Different choices of partition
functions change the matching densities to our mean field
results slightly, but the influence on the thermodynamics is
negligible. This is also discussed in Ref.~\cite{NSE09}.

Chemical equilibrium between nucleons and a heavy nucleus
with $Z$ protons and $N$ neutrons insures, 
\beq 
\mu_i\ =\ Z\mu_p\ +\ N\mu_n, 
\eeq 
where $\mu_i, \mu_p, \mu_n$ are chemical potentials
of the heavy nucleus, protons and neutrons, respectively. Therefore the
fugacity of a heavy nucleus is readily obtained,
\beq\label{ion_fuga} 
z_i\ =\ \mathrm{exp}(\mu_i+E_i)/T\ =\ z_p^Zz_n^N e^{E_i/T}, 
\eeq where $E_i$ is the binding energy of
heavy nucleus $^A Z$.

We consider the Coulomb interaction between the electron background and
nucleus following Baym \etal \cite{BBP}, but generalized to multiple species
of nuclei.  The total Coulomb energy of a nucleus in an electron
background is, 
\beq\label{QC} 
Q_i^C\ =\ \frac 3 5
\frac{Z_i^2\alpha}{r_A}[1-\frac 3 2 \frac {r_A} {r_i} + \frac 1 2
(\frac {r_A} {r_i})^3], 
\eeq 
where $r_A = 1.16 A^{1/3}$ fm is the
nuclear radius (in accordance with that in FRDM \cite{FRDM}), and
$r_i$ is the average ion sphere radius defined through 
\beq \frac
4 3 \pi r_i^3 (\sum_j Z_j n_j)\ =\ Z_i. 
\eeq 
We emphasize that the sum over $j$ runs over heavy nuclei.  This definition ensures local
charge neutrality inside each ion sphere.  The first term in Eq.~(\ref{QC}) comes from the proton-proton Coulomb energy in the nucleus and is also included in the binding energy of the nucleus. When $r_A$ approaches $r_i$ the total Coulomb energy in a heavy nucleus vanishes as expected.  So the Coulomb correction to the binding energy of nucleus is 
\beq\label{coulomb1}
E_i^C\ =\ \frac 3 5 \frac{Z_i^2\alpha}{r_A} \bigl[-\frac 3 2
\frac {r_A} {r_i} + \frac 1 2 (\frac {r_A} {r_i})^3\bigr]. 
\eeq 
Adding this correction to the binding energy of a nucleus,
Eq.~(\ref{ion_fuga}) becomes
\beq\label{ion_fuga2} 
z_i\ =\ \mathrm{exp}(\mu_i+E_i-E_i^C)/T\ =\ z_p^Zz_n^N e^{(E_i-E_i^C)/T}.
\eeq
The densities of each species can be obtained from
\beq n_a = z_a
\left( \frac{\partial}{\partial z_a} \frac{\mathrm{log}{\cal
Q}}{V}\right)_{V,T}.\eeq
This gives,
 \beqn\label{densityn}  n_n\ &=&\ \frac 2
{\lambda_n^3}[z_n+2z_n^2b_n+2z_pz_nb_{pn}+8z_\alpha z_nb_{\alpha
n}],\\ \label{densityp} n_p\ &=&\ \frac 2
{\lambda_n^3}[z_p+2z_p^2b_n+2z_pz_nb_{pn}+8z_\alpha z_p b_{\alpha
n}], \\
 n_\alpha\ &=&\ \frac 1
{\lambda_\alpha^3}[z_\alpha + 2z_\alpha^2 b_\alpha+ 2z_\alpha(z_n+z_p)b_{\alpha n}],\\
\label{density_i} n_i\ &=&\ \frac 1 {\lambda_i^3}z_i\Omega_i.
\eeqn 
The mass fraction of species $a(A,Z)$ is defined as,
\beq\label{massfraction} X_a\ =\ A_a n_a /n_B. \eeq
The two conditions used to determine the fugacities of the neutrons and
protons are that the total baryon density is conserved,
\beq\label{condition1} n_B\ =\ n_n + n_p + 4n_\alpha +
\sum_{\substack{i}} A_i n_i, 
\eeq 
and that the proton fraction is given by,
\beq\label{condition2} Y_P\ =\ (n_p + 2n_\alpha +
\sum_{\substack{i}} Z_i n_i)/n_B. 
\eeq 
Since the Coulomb correction is included, one extra loop is needed to self-consistently determine the values of the ion sphere radii $r_i$ for each species.

The entropy density $s$ of the Virial gas is obtained from,
\beqn\label{entropy1}
 s &&= \left( \frac{\partial P}{\partial T}\right)_{\mu} =\ \frac 5
2 \frac P T\ + \frac{2T}{\lambda_n^3}[(z_n^2+z_p^2)b'_n+2z_nz_p
b'_{pn}] \cr && +\
\frac{T}{\lambda_\alpha^3}[z_\alpha^2b'_\alpha+2z_\alpha(z_n+z_p)b'_{\alpha
n}]\cr && \  - \left[\sum_{\substack{i}}
n_i\mathrm{log}z_i+n_n\mathrm{log}z_n+n_p\mathrm{log}z_p+n_\alpha\mathrm{log}z_\alpha\right]
\cr &&+ \sum_i\frac{z_i}{\lambda_i^3}T\Omega'_i -\sum_i n_i
\frac{\partial E_i^C}{\partial T}|_\mu, 
\eeqn 
where the prime indicates a derivative with respect to temperature. Note that the last term is
the Coulomb correction to the entropy, which is nontrivial to
evaluate directly.   However the free energy of the Virial gas can be
obtained directly (see discussion below) so that we can evaluate the
entropy from the derivative of the free energy.

The energy density $\epsilon$ can be calculated from the entropy density,
\beq\label{energy1} \epsilon\ =\ Ts + \sum_{\substack{a}}n_a \mu_a
- P, 
\eeq
where the sum on $a$ runs over neutrons, protons, alphas and
heavy nuclei.

The free energy density $f$ is given by 
\beqn
f\ &=&\
\epsilon - Ts\ =\ \sum_{\substack{a}}n_a \mu_a - P\cr &=&\ n_n
T\log z_n\ +\ n_p T\log z_p + n_\alpha T\log z_\alpha - n_\alpha
E_\alpha \cr && + \sum_i \left[ n_i T\log z_i - n_i(E_i -
E_i^C)\right] - P. \eeqn From the above equation we define an
effective Coulomb correction to free energy per nucleon $\Delta f/A$,
\beq\label{coulombv}
\Delta f/A\ =\ \frac{\sum_i E_i^C n_i}{n_B}. 
\eeq 
The free energy per nucleon is 
\beq \label{fe:Virial} 
F/A = f/n_B. 
\eeq

The thermodynamic pressure $P$ can be obtained from the free
energy, 
\beq\label{pressure:v1} 
P_{th}\ =\ n_B^2
\left(\frac{\partial (F/A)}{\partial n_B}\right)_{T,Y_P}, \eeq which
can be rewritten as \beq\label{pressure:v2} P_{th}\ =\ P +
n_B\sum_i n_i\frac{\partial E_i^C}{\partial n_B}|_{T,Y_P}. 
\eeq
The second term in Eq.~(\ref{pressure:v2}) is the correction to
thermodynamic pressure due to the Coulomb interaction.  We present results for the Virial equation of state in Section \ref{result}.

\section{\label{numerics}Numerical details}

In Subsection \ref{partition} we describe some details of the evaluation of the partition functions and then in Subsection \ref{computation} we briefly describe the parallel computations of the EOS for many different temperatures, densities, and proton fractions. 

\subsection{\label{partition}Recipes for partition function}

Fowler \etal \cite{Fowler78} proposed an efficient approximation
for the partition function of hot nuclei,
which has a closed form, 
\beq\label{partition2}
\Omega_i\ =\ \Omega_d + \int_{E_d}^{E_t}
dE\rho(E)\mathrm{exp}(-E/T) - \Omega_c, 
\eeq 
where $\Omega_d$ is the contribution from known discrete states and $\Omega_c$ is the
continuum subtraction.  One could use $\Omega_d = (2J_0+1)$ where
$J_0$ is the ground state spin.  Inaccuracies from this approximation
will become progressively unimportant beyond 10$^{10}$ K. The
continuum subtraction, on the other hand, only becomes important
beyond 10$^{11}$ K (see also \cite{Nadyozhin04}), by this
temperature uniform matter will be more stable in most regions of phase
space, as will be shown in the following discussions. So in the
temperature range we are interested, one can discard the latter
term.  Therefore the nuclear partition function becomes,
\beq\label{partition3} 
\Omega_i\ =\ (2J_0+1) + \int_{E_d}^{E_t}
dE\rho(E)\mathrm{exp}(-E/T), 
\eeq 
and its derivative versus temperature is, 
\beq\label{deri}T \Omega'_i\ =\
 \int_{E_d}^{E_t} dE\rho(E)\mathrm{exp}(-E/T)\frac E T. \eeq
A widely used expression for the level density $\rho(E)$ is the back-shifted
Fermi gas formula \cite{Bethe36}, 
\beq\label{bsfg} 
\rho(E)\ =\
\frac{\sqrt{\pi}}{12}\frac{\mathrm{exp}(2\sqrt{aU})}{a^{1/4}U^{5/4}},
\eeq 
where $a$ is the level spacing, and $U = E - \delta$ with
$\delta$ a backshift parameter related to pairing.  Various
prescriptions of these two parameters are available, which
reproduce more rigorous results for the level density from Monte Carlo or combinatorial calculations. We will use the prescription from Fowler et
al \cite{Fowler78} in our calculation and include that of Rauscher
\etal \cite{Rauscher97} as an option in our code.

The integral limits are determined by the following equations
\cite{Fowler78}: 
\beqn E_d &=& \frac 1 2 \mathrm{min}(S_n, S_p),
\\ E_t &=& \mathrm{min}(S_n+E_R, S_p+E_R+\frac 1 2 E_c). 
\eeqn
$S_n$ and $S_p$ are the single neutron and single proton separation
energies, which could be obtained from the mass table. $E_R$ is
the zero-point energy,  $E_R=\hbar^2/2MR^2$ with
$R=1.25(A-1)^{1/3}$.  The Coulomb barrier is $E_c=(Z-1)\alpha/R$.
For unstable nuclei ($S_n$ or $S_p$ $<$ 0), the partition function
is set to the ground state value $\Omega_d$. There is still a large
ambiguity regarding the value of the integral upper limit $E_t$,
which is related to the contribution of the continuum. This ambiguity
will introduce big changes in the partition function
only when the temperature is above several MeV (few times
10$^{10} K$).  We compared results with different choices of
$E_t$ and found the ambiguity changes the transition density to our relativistic mean
field EOS \cite{SHT10a} slightly, but the influence on the thermodynamics is
negligible. This was also discussed in a previous work \cite{NSE09}.

When $\delta$ is larger than $E_d$, $\delta=E_d$ should be used.
In this case Eq.~(\ref{bsfg}) is inapplicable due to the pole in
the lower integration limit.  A constant temperature formula,
\beq\label{ct} 
\rho(E)\ =\ C \mathrm{exp}(U/T_c), 
\eeq 
is used instead.  The constants $C$ and $T_c$ are obtained by the
continuity of the level density and its first derivative when matched
to Eq.~(\ref{bsfg}).

\subsubsection{Choices of a and $\delta$ by Fowler \etal.
\cite{Fowler78}}

In our calculation we use the prescription of Fowler \etal
\cite{Fowler78} for the level spacing parameter $a$ (in MeV$^{-1}$)
and pairing parameter $\delta$ (in MeV): 
\beq
\begin{array}{c} Z\leq 30: \ a\ =\ 0.052A^{1.2},\ \delta = \delta_p - 80/A,\\
  Z\geq 30:\ a\ =\ 0.125A,\ \delta = \delta_p - 80/A - 0.5,\\
\end{array} \eeq where $\delta_p = (11\cdot A^{-1/2} \mathrm{MeV})
\left[1+\frac 1 2 (-1)^Z + \frac 1 2 (-1)^N\right]$.
When $\delta > E_d$, $\delta$ is set to $E_d$, and Eq.~(\ref{ct})
is used below an energy 2$U_c\ (E_t>2U_c>0)$.  By continuity
of the level density and its derivative, the constants $C$ and $T_c$
are obtained, 
\beqn 
\frac 1 {T_c}\ &=&\ -\frac 5 4 \frac 1
{U_c} + \frac {\sqrt{a}} {\sqrt{U_c}}, \cr C\ &=&\
\frac{\sqrt{\pi}}{12}a^{-1/4} U_c^{-5/4} \mathrm{exp}(\frac 5
4+\sqrt{aU_c}). 
\eeqn 
In our calculation $U_c$ is chosen to be $\delta$.

\subsubsection{Choices of a and $\delta$ by Rauscher \etal .
\cite{Rauscher97}}

In the parameterization of Rauscher \etal , the level spacing $a$ depends on energy, 
\beq 
a(U,Z,N)\ =\ (\alpha A+ \beta
A^{2/3}) [1 + b\cdot\frac{1-e^{-\gamma U}}{U}], 
\eeq
where $b=E_{mic}$ describes properties of a nucleus differing from the
spherical macroscopic energy, which is already given in the FRDM mass
table. Based on the FRDM mass formula, the best fitted values for
$\alpha, \beta$, and $\gamma$ are obtained by comparing the level
densities with experimental analysis: $\alpha$ = 0.1337, $\beta$ =
-0.06571, $\gamma$ = 0.04884 \cite{Rauscher97}.

The pairing parameter $\delta$ is given by, 
\beq 
\delta = \frac 1 2
\{\Delta_n(Z,N)+\Delta_p(Z,N)\}, 
\eeq 
where the neutron pairing energy is 
\beq 
\Delta_n(Z,N)\ =\ \frac 1 2
[2E^G(Z,N)-E^G(Z,N-1)-E^G(Z,N+1)], 
\eeq 
and $E^G(Z,N)$ is the binding energy of nucleus $(Z,N)$.  The proton
pairing energy $\Delta_p(Z,N)$ is calculated in a similar way.  The constants $C$ and $T_c$ are obtained as following, 
\beqn 
\frac 1 {T_c}\ &=&\ -\frac 5 4 \frac 1 {U_c}+\sqrt{\frac{a}{U_c}}-\frac
1 4 \frac {a'} a+ \sqrt{\frac{U_c}{a}} a', \cr C\ &=&\ \rho(U_c)
e^{-U_c/T_c}, \eeqn where \beq a'\ =\ (\alpha A+ \beta A^{2/3})
\frac b {U_c^2} (-1+e^{-\gamma U_c}+\gamma U_c e^{-\gamma U_c}).
\eeq

\subsection{\label{computation}Computational methodology}

We calculate, in parallel, the points in parameter space covered by the Virial EOS, analogous to the way used in our previous paper to perform relativistic mean field calculations\cite{SHT10a}.   There are a total of 73,840 points in the 3-dimensional parameter space of  temperature $T$, proton fraction $Y_P$, and density $n_B$.  Each point takes up to 20 minutes to calculate.  The overall Virial EOS took about 1,000 CPU days in Indiana University's supercomputer cluster.

Each point in the parameter space was mapped to a unique integer that we 
refer to as the job index.  A file, {\it runlist}, was prepared with a 
list of job indicies for the whole parameter space, and a single character 
(A=available, R=running, r=Re-running,
C=complete, T=time-limited and F=failed) that gives the status of 
calculations for that job index.  A Message Passing Interface (MPI) parallel wrapper code manages 
the running of the many requested  tasks.  Typically, one parallel job requests a set of compute cores (usually 256). Each MPI rank,
using a single CPU core, is assigned one job index 
corresponding to one point in the parameter space and evaluates 
the required quantities.

Initially, rank zero of the MPI job 
\begin{itemize}
\item{locks the job listing file {\it runlist}},
\item{reads {\it runlist} until a list of available 
tasks is filled},
\item{closes {\it runlist} and releases the lock},
\item{passes a job index to each MPI rank and begins the calculation
for that job index.}
\end{itemize}

When the calculation completes (or time-limits or fails) for a given 
MPI rank, the status character for the job index in 
{\it runlist} is modified appropriately. The now available 
MPI rank will search {\it runlist} for the next available task and the 
calculation restarts for the new job index.   Since completion occurs asynchronously file locking is not used for this part of the process.

A simple batch job runs through the points in parameter space. A wall clock 
limit (48 hours) is used. Each rank of the MPI job can run a series
of points using the above procedure, efficiently using each available core for 
the requested wall clock period.  One job per core is running when the wall clock 
limit is reached. These jobs are identified by being left in the "R" state after the batch job completes.  This procedure allows us to calculate $>$ 99 $\%$ of the points in the {\it runlist} file.

\section{\label{result}Results}

In this section we discuss the Virial EOS in detail.  First we show the effect of different mass tables on the EOS and its influence on the matching to RMF results.  We also discuss the coulomb correction and compare our results with some analytical cluster expansion analysis. Second we discuss the matching between the Virial EOS and RMF EOS for several choices of temperature and proton fraction.  Finally we show some examples of the distribution of nuclei obtained in the Virial EOS. We also compare some examples of mass fractions of nuclear matter between Virial EOS and existing EOS tables.

\subsection{Effect of different mass tables}

We use both the FRDM \cite{FRDM} and HFB14 \cite{HFB14} mass tables in the Virial expansion.  Here we compare the free energy and average charge number $Z$ (average mass number $A$ = $Z$/$Y_P$) from the Virial expansion with the two different mass tables. We also match them to the relativistic mean field results from our previous paper \cite{SHT10a}. 

\subsubsection{Free energy}

\begin{widetext}

\begin{figure}[htbp]
 \centering
 \includegraphics[height=8.5cm,angle=-90]{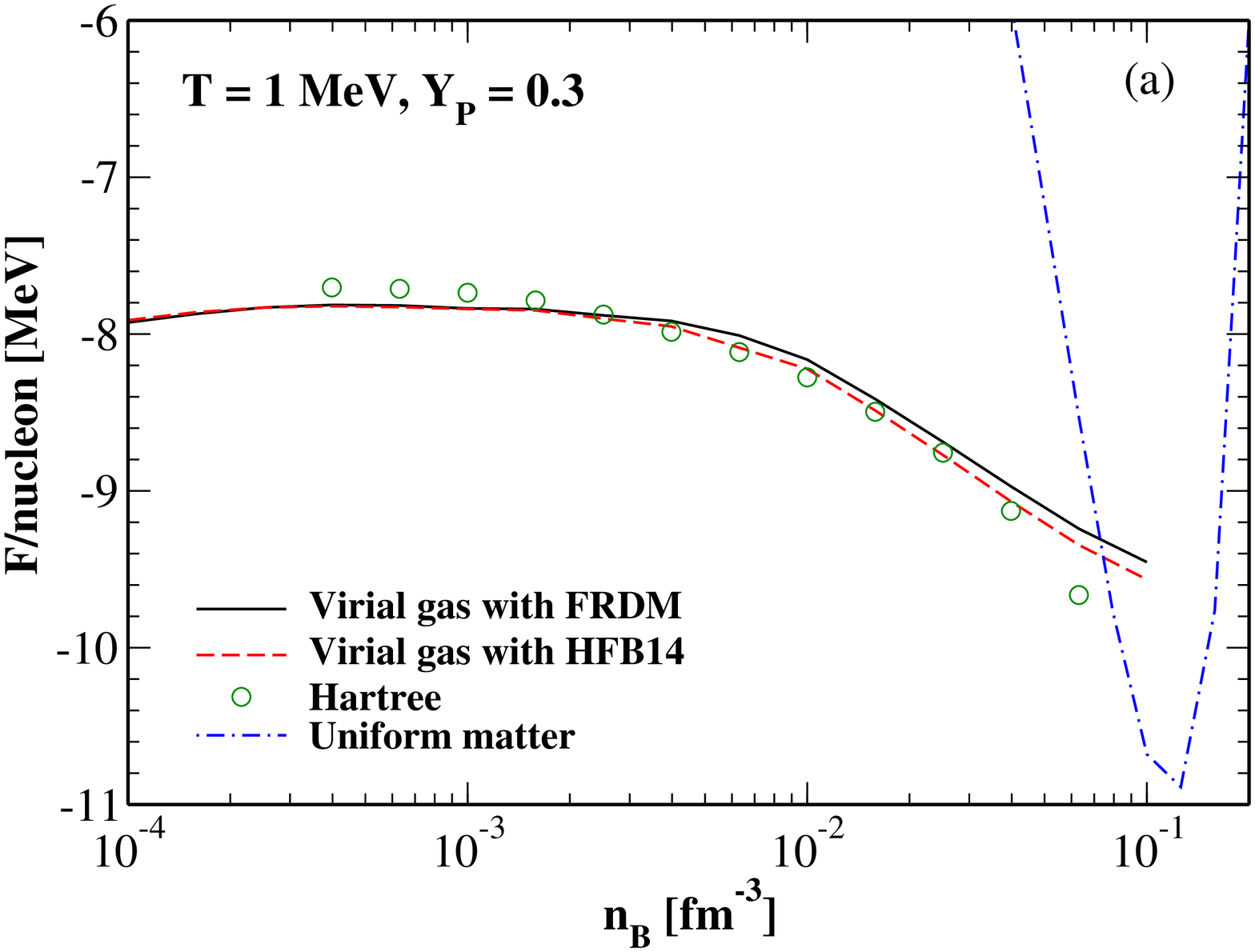}
 \includegraphics[height=8.5cm,angle=-90]{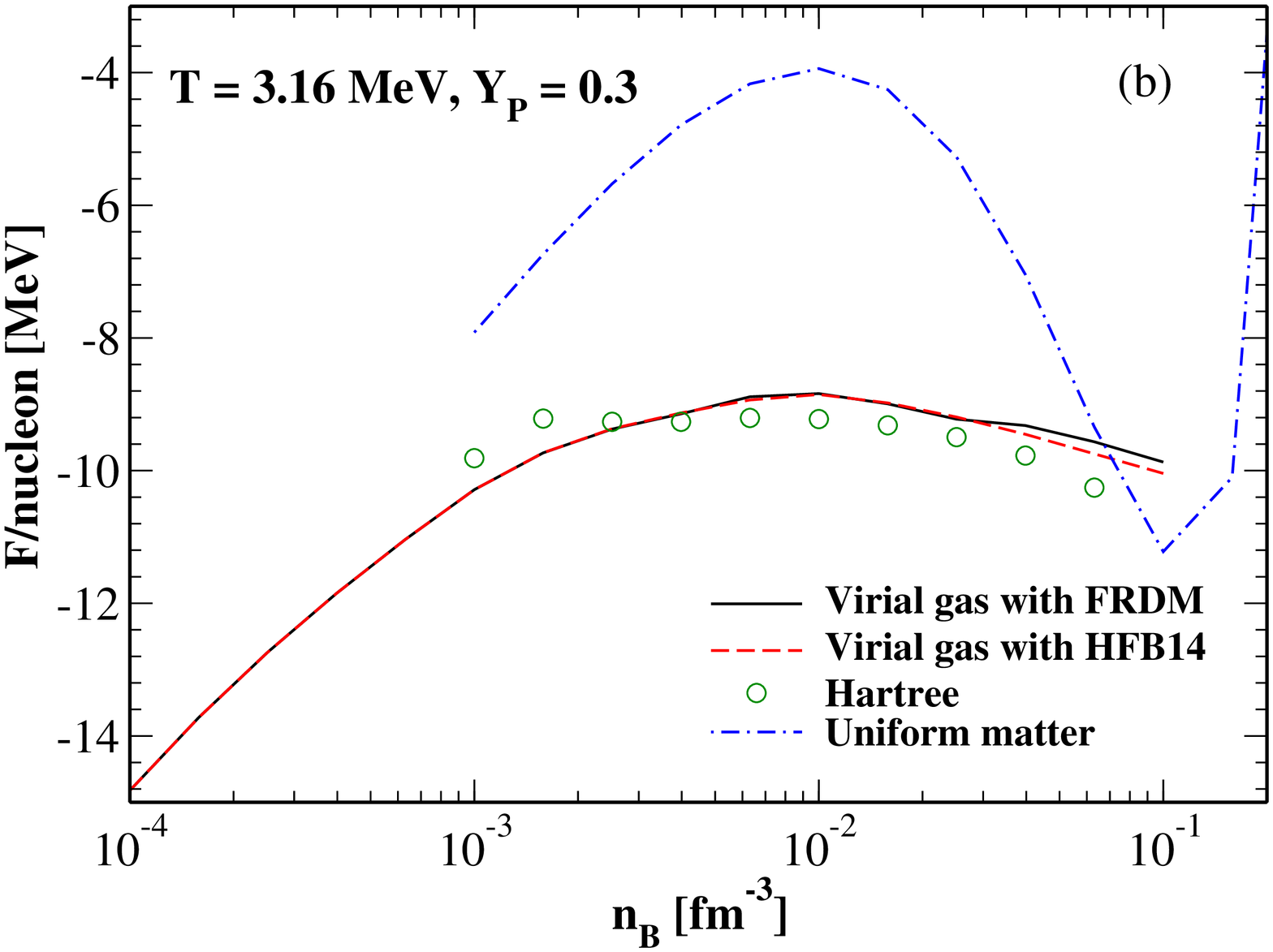}

\caption{(color on line) Free energy per nucleon of nuclear matter at $T$ = 1 (a) and 3.16 MeV (b) with $Y_P$ = 0.3.}\label{fig:fe_masstable}
\end{figure}

\end{widetext}

In Figure \ref{fig:fe_masstable}, free energies are shown as a function of density for nuclear matter at $T$ = 1 and 3.16 MeV and $Y_P$ = 0.3.  Virial gas results are obtained from Eq.~(\ref{fe:Virial}) for the FRDM (black solid curve) and HFB14 (red dashed curve) mass tables.   Nonuniform (or Hartree) mean field calculations (circles) and uniform matter calculations (dotted dashed curve) are also shown (See Eqs. (10) and (18) in Ref.~\cite{SHT10a}).  These calculations give lower free energies at high densities. The two different mass tables in the Virial expansion give very similar free energies for nuclear matter.  The difference between the black curve (FRDM) and red curve (HFB14) is very small until densities where the Hartree mean field results are more stable than either mass formula. For these two cases, the transition from Virial to Hartree EOS occurs at a density of about 3.98$\times$10$^{-3}$ fm$^{-3}$.

\subsubsection{Average $A$ and $Z$}

\begin{widetext}

\begin{figure}[htbp]
 \centering
 \includegraphics[height=8.5cm,angle=-90]{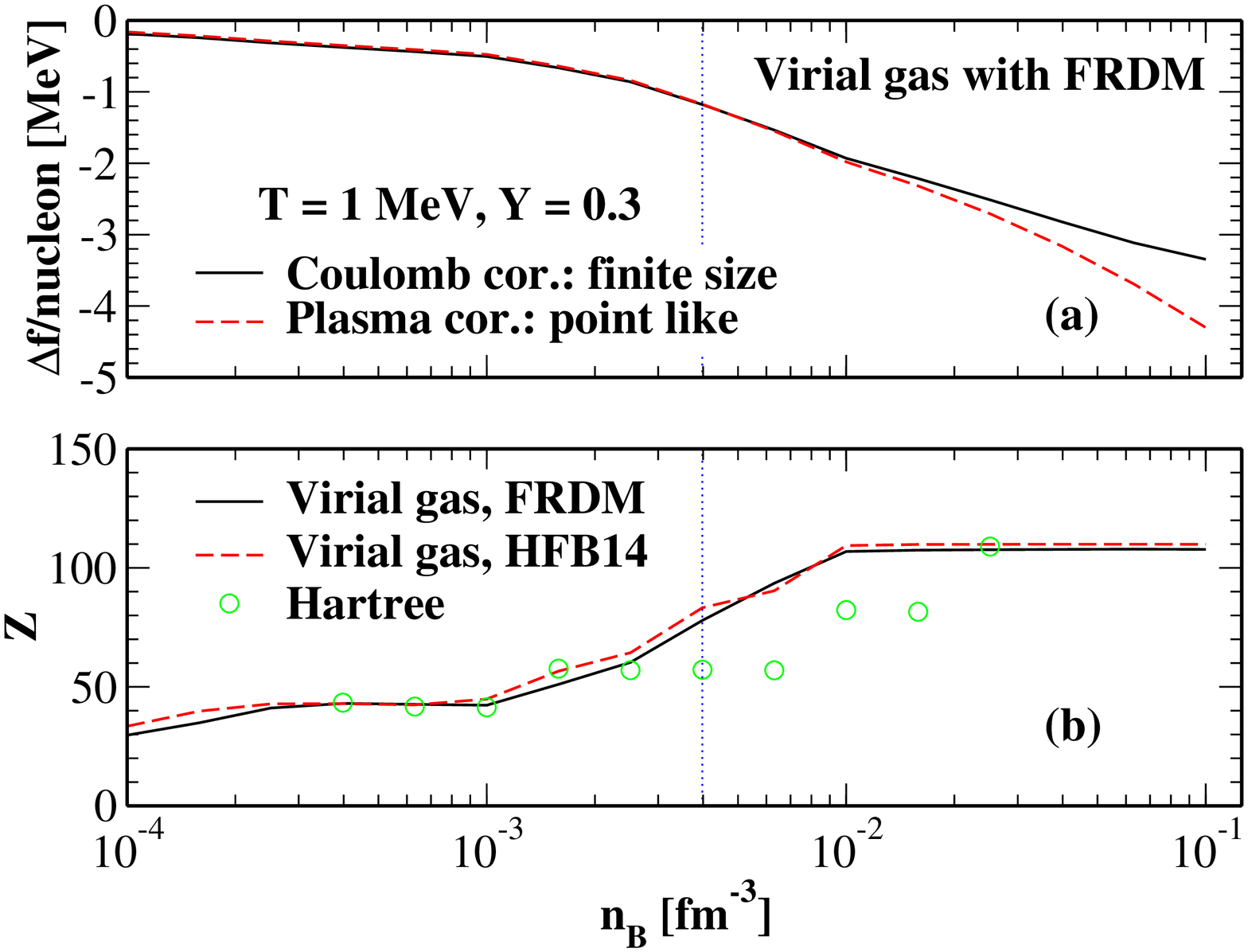}
 \includegraphics[height=8.5cm,angle=-90]{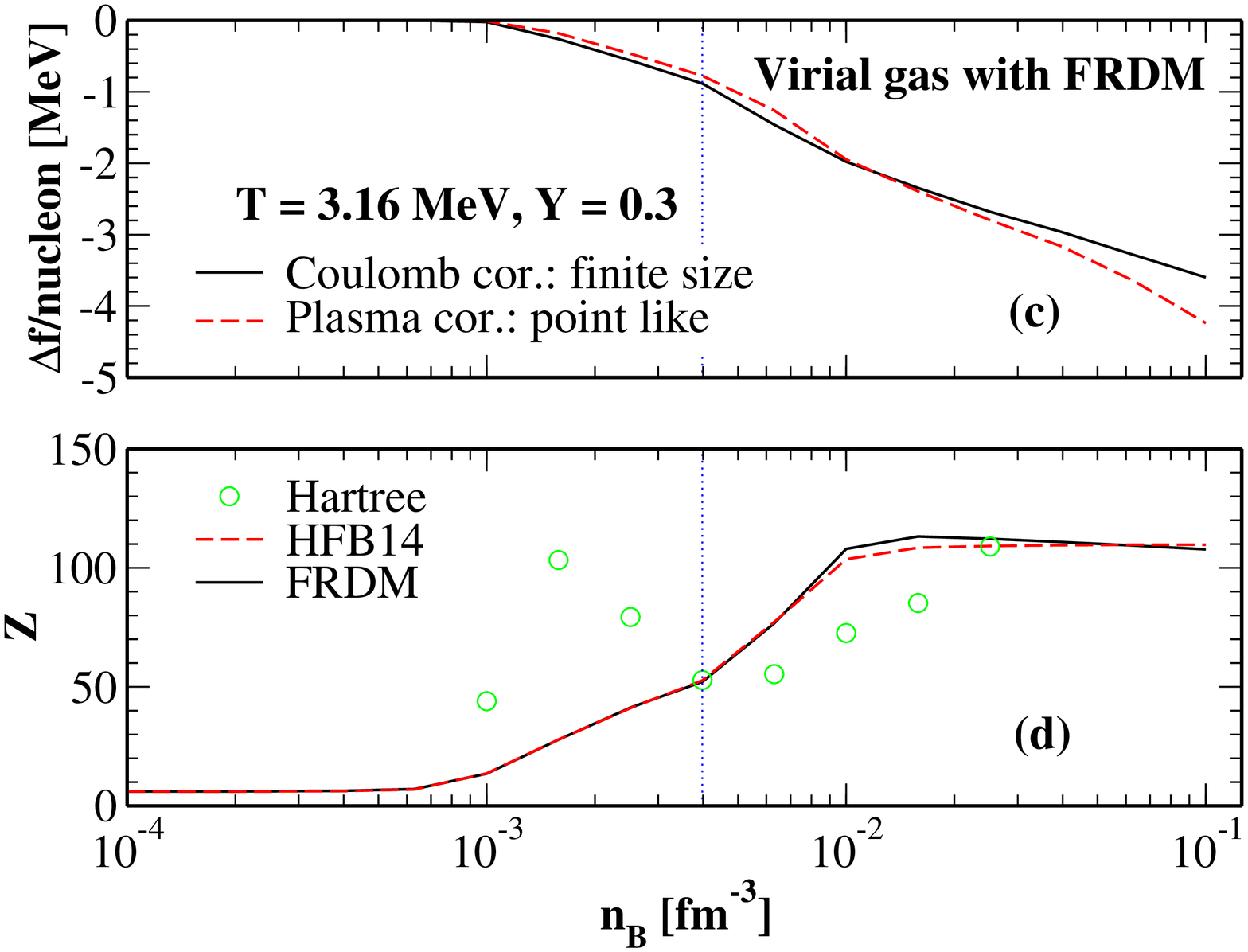}

\caption{(color on line) Upper panels (a), (c) show Coulomb corrections; lower pannels (b), (d), show average charge number of heavy nuclei, in nuclear matter at $T$ = 1 MeV, left panels (a), (b), and $T$=3.16 MeV, right panels (c), (d).  The proton fraction is $Y_P$ = 0.3.}\label{fig:cou_masstable}
\end{figure}

\end{widetext}

Similar to Fig. \ref{fig:fe_masstable}, the lower panels in Fig. \ref{fig:cou_masstable} give the corresponding average charge number $Z$ from the Virial EOS with either FRDM or HFB14 mass tables, and from Hartree mean field results.  Again the Virial EOS with the two mass tables predict very similar values for $Z$.  Moreover, the Virial EOS with either mass table gives very similar $Z$ to that from the Hartree mean field results at the transition density 
3.98$\times$10$^{-3}$ fm$^{-3}$ (blue dotted line). The fluctuation of $Z$ in Hartree results below the transition density (at $T$ = 3.16 MeV) is probably due to finite step error in search of cell size. In conclusion, the composition, free energy, and transition density to Hartree mean field results, depend little on the choice of mass tables.

In the two upper panels in Fig. \ref{fig:cou_masstable}, we compare the Coulomb energy correction in our Virial expansion, Eq.~(\ref{coulombv}), with an analytic cluster expansion for the one-component plasma, Eq. (22) in \cite{coulomb}. The overall agreement for densities below the Virial-Hartree transition is good, though at higher temperature the differences become larger. The reason is probably that we calculate the multi-component contribution to the Coulomb correction in the Virial gas while we only used the average charge number in the analytic formula for the one-component plasma.

\subsection{Free energy and phase boundaries}

\begin{widetext}

\begin{figure}[htbp]
 \centering
 \includegraphics[height=8.5cm,angle=-90]{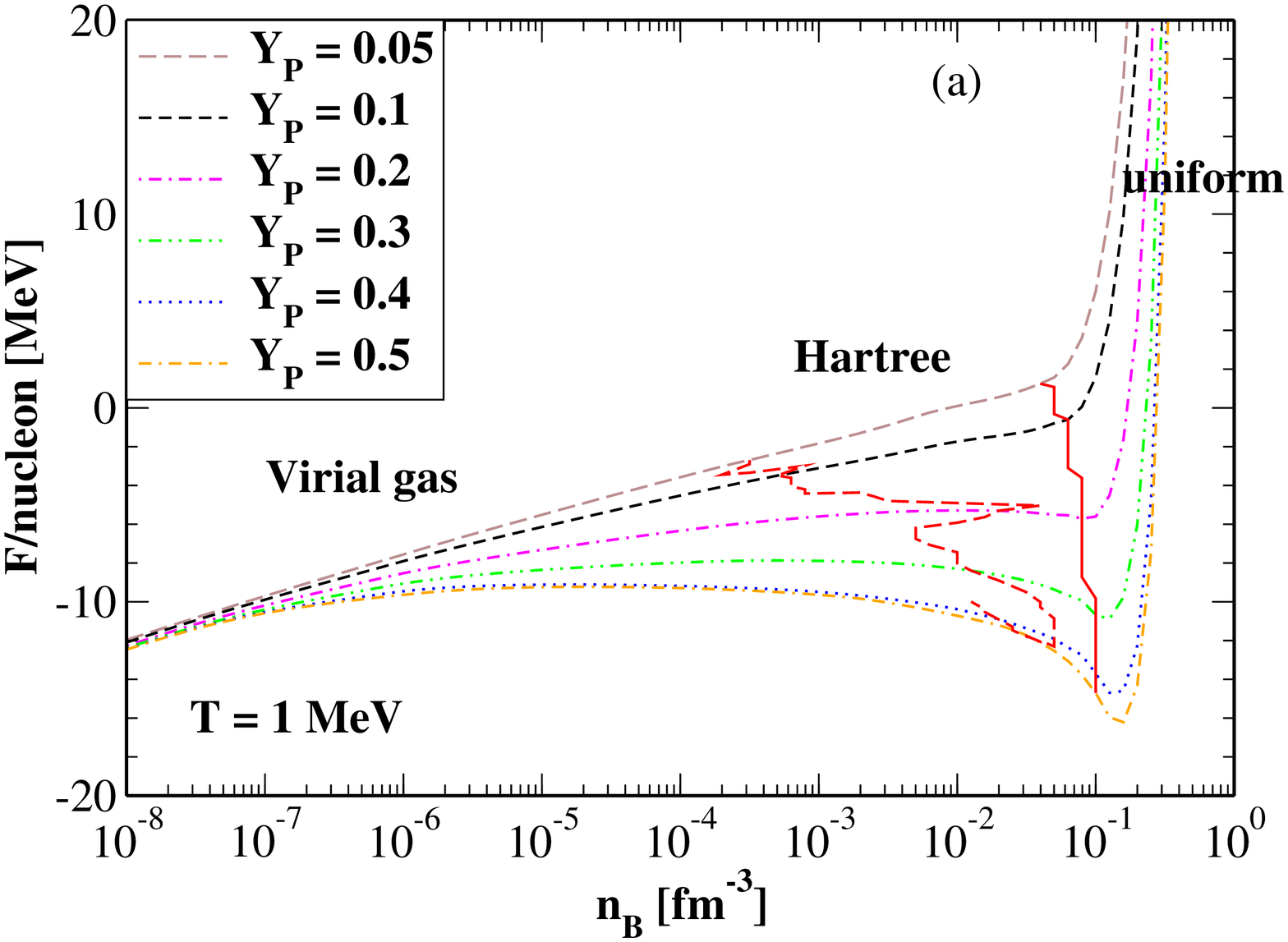}
 \includegraphics[height=8.5cm,angle=-90]{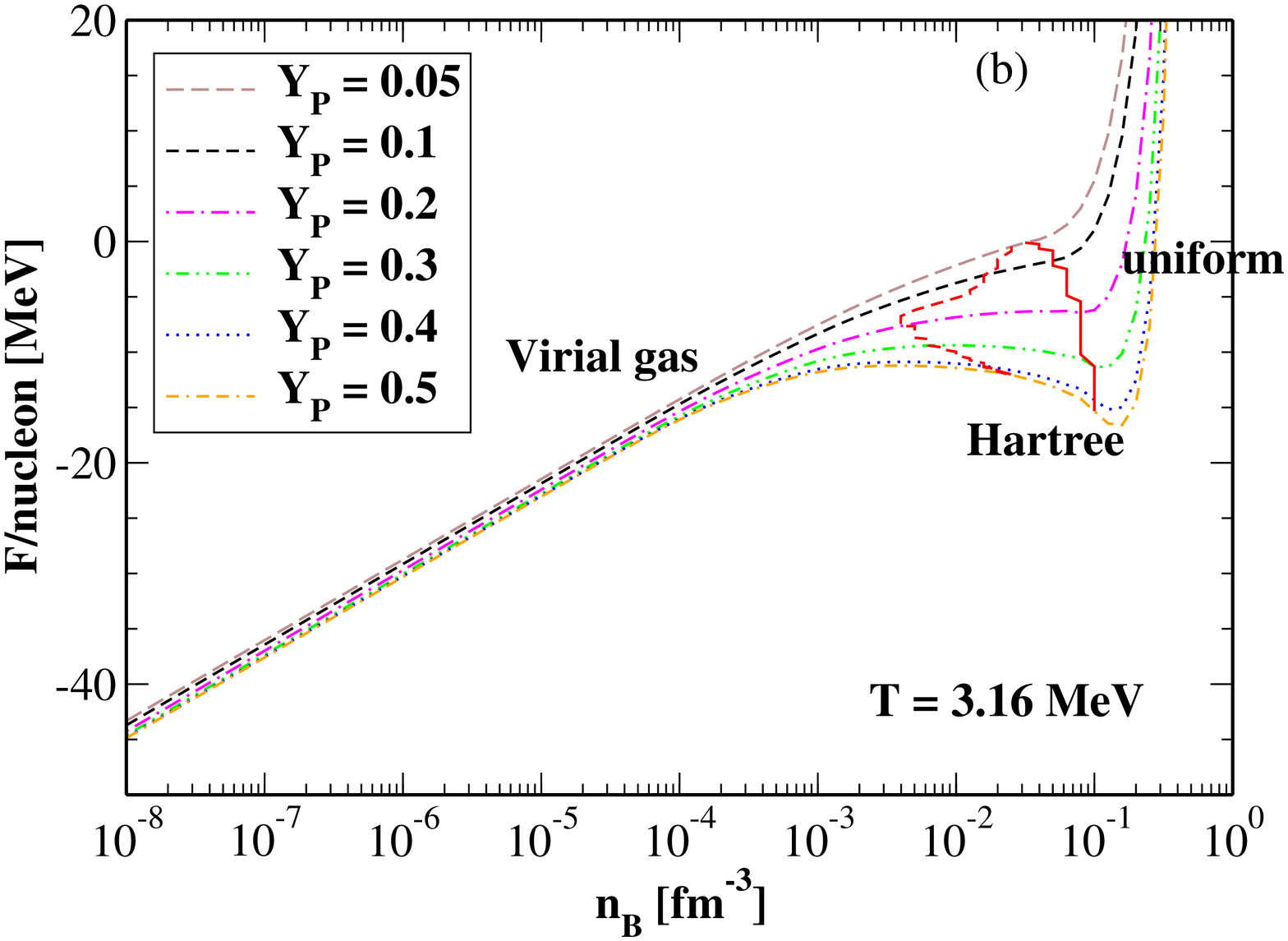}
 \includegraphics[height=8.5cm,angle=-90]{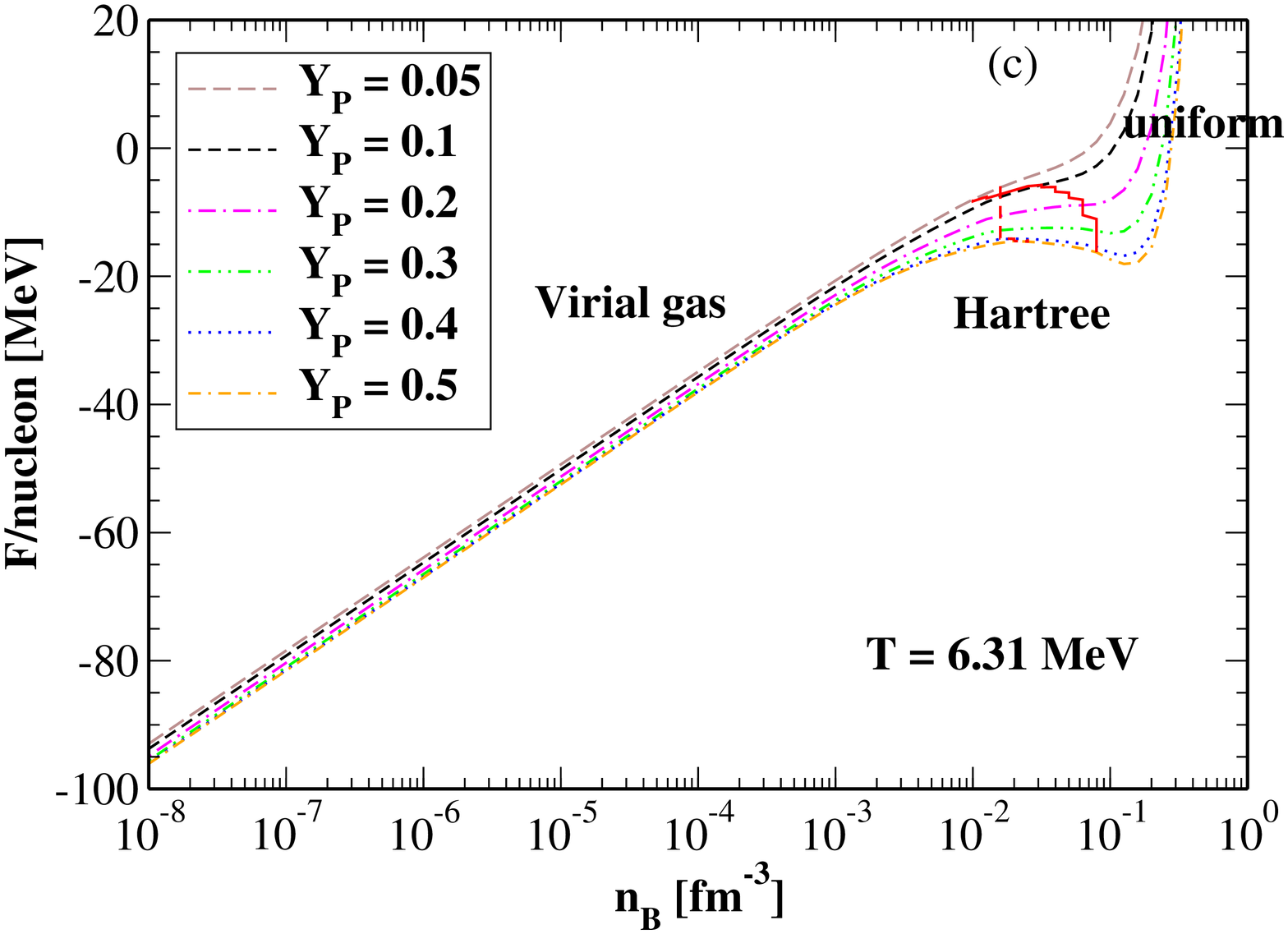}
 \includegraphics[height=8.5cm,angle=-90]{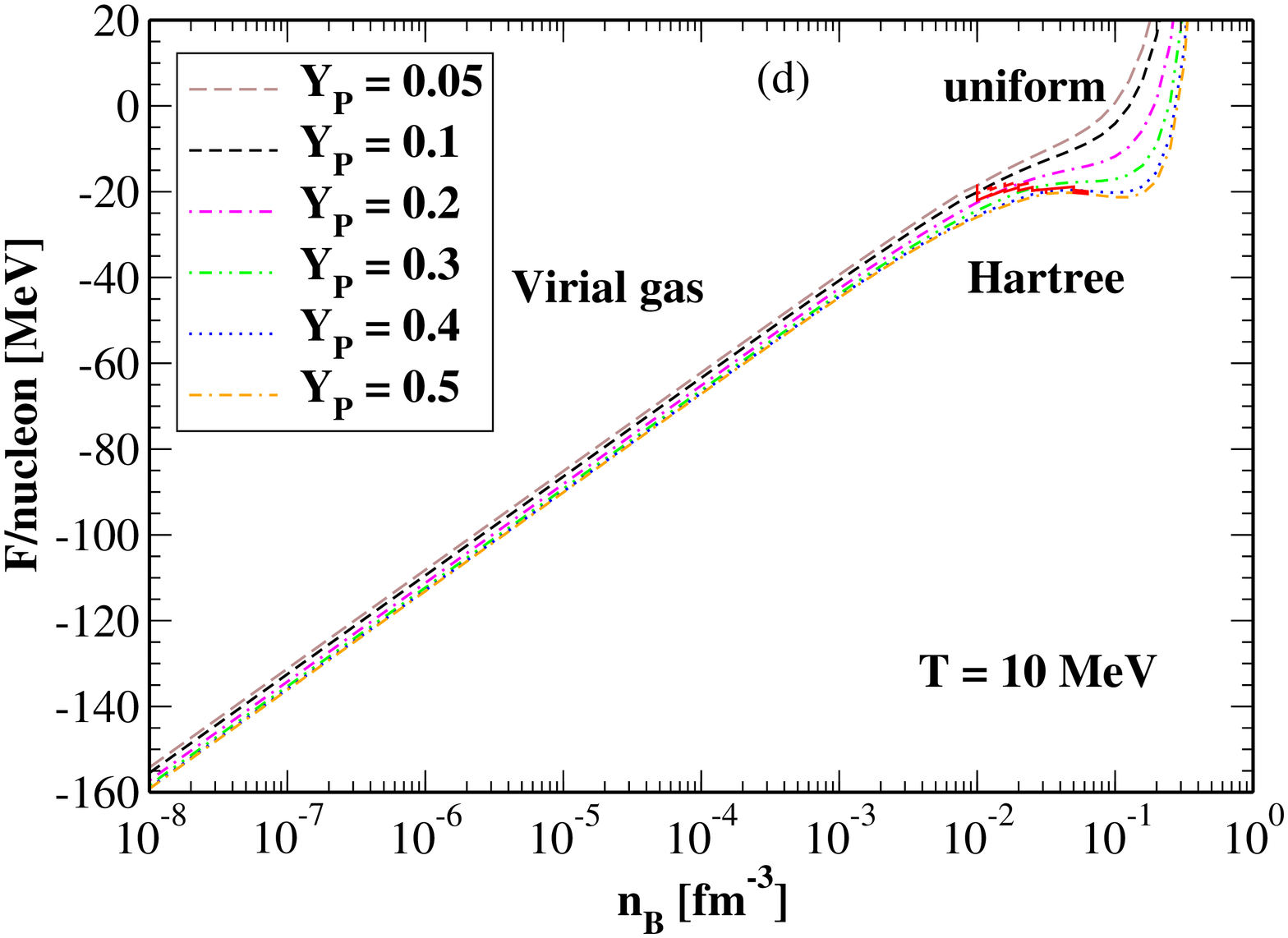}

\caption{(color on line) Free energy per nucleon of nuclear matter at
temperatures of $T=1$ (a), 3.16 (b), 6.31 (c), and 10 (d) MeV.  The proton fraction ranges from $Y_p=0.05$ to 0.5. }\label{fig:fe}
\end{figure}

\end{widetext}

We show in Fig.~\ref{fig:fe} the transition densities between the Virial and RMF EOSs.  We also show the free energy per nucleon $F/A$ as a function of density $n_B$ for $T$ = 1, 3.16, 6.31 and 10 MeV.   At low densities, $F/A$ is obtained from Eq.~(\ref{fe:Virial}) in
the Virial expansion.  The free energy per nucleon $F/A$ is also shown for Hartree mean field calculations at intermediate densities, and from uniform matter at high densities. The latter two have been obtained in our first paper \cite{SHT10a}.

In most cases the transition (as density grows) is found at the density when the
Hartree or uniform matter calculation gives a lower free energy compared to Virial results. For matter at very low temperature (not greater than $\sim$ 1 MeV) and
low proton fraction (not greater than $\sim$ 0.1), some matching points
are obtained at the density when the Virial gas and Hartree calculation
give the closest free energy. The difference is below hundreds of
KeV and is comparable to the mean deviation of binding energy for nuclei ($\sim$ 600 KeV) in the FRDM mass table \cite{FRDM}.

In each panel in Fig.~\ref{fig:fe}, the vertical red dashed curves give the Virial gas -
Hartree Wigner-Seitz cell transition densities, and the red solid curves give
the transition densities to uniform matter.  As temperature
increases, the second transiton may happen at lower density than that
of first transiton, which means the Hartree Wigner-Seitz cell is energetically
unfavorable for all densities.  This indicates the critical temperature
for the nucleon liquid - Virial gas transition.  We note here that the Virial
gas may still contain a small amount of alpha particles and heavier
nuclei even above this transition temperature.   As seen from the figures, for matter at any temperature and proton fraction the Virial - mean field transition densities
are larger than a few times 10$^{-4}$ fm$^{-3}$, which is about the
neutrinosphere density.  So in almost all regions around and
below the neutrinosphere density, our EOS is represented by the Virial
results, which include multiple nuclei.

For matter at low density in the Virial gas phase, the free energy
scales nicely with density. The reason is as follows. Low density
matter is dominated by neutrons and protons, and the interactions
among them are not important due to the large particle spacing, so the
kinetic energy dominates and scales as the logarithmic value of
the density.  As the temperature rises, the scaling behavior extends to
higher densities until heavy nuclei appear, and
electromagnetic and strong interaction corrections become
important.  Formation of nuclei greatly decreases the free energy, both
in the Virial gas and Hartree mean field regimes.

\subsection{Mass fractions of species}

The Virial expansion gives the distribution of heavy nuclei 
(refer to Eq.(\ref{density_i}, \ref{massfraction})), where 8980
species of nuclei are in thermal and chemical equilibrium with
free neutrons, free protons and alpha particles. This is an improvement over the L-S EOS and S-S EOS that both used a single-nucleus representation.

Figure~\ref{fig:dist_T1Y2n-4} shows mass fractions of different
nuclei ($Z, A$) for matter with $T$ = 1 MeV, $n_B$ = $10^{-4}$ fm$^{-3}$, and $Y_P$ = 0.2. In upper panel different colors indicate the mass fraction using a $ \rm{Log}_{10} $ scale.  In lower panel different lines indicate the contours of mass fraction in $ \rm{Log}_{10} $ scale from -1 to -7. The total mass fraction of heavy nuclei is 56.5\% (the rest is free neutrons). The majority of the mass fractions are centered around $Z$ = 25 $\sim$ 30, with a wide range of other nuclei with smaller abundances.  In Figure~\ref{fig:dist_T1Y4n-4} for a higher $Y_P$=0.4, the total mass fraction of heavy nuclei is 99.1\% (free neutrons and protons have 0.8\% and 0.1\% respectively).  The majority mass fractions are centered around $Z$ = 30 $\sim$ 35.

In Fig.~\ref{fig:dist_T1Y4n-3} we show mass fractions of different nuclei for matter with $T$ = 1 MeV, $Y_P$ = 0.4, and $n_B$ = $10^{-3}$ fm$^{-3}$.  The total mass fraction of heavy nuclei is near one and the distribution is centered around $Z$ = 35 and 50. The mass distribution of heavy nuclei in this high $Y_P$ case is a double-peaked Gaussian distribution, as shown in Fig.~\ref{fig:z(n)}, where $n(Z)$ is sum of the abundances of heavy nuclei with same proton number $Z$.
 At a higher $T$ = 3.16 MeV shown in Fig.~\ref{fig:dist_T3Y4n-3}, the total mass fraction of heavy nuclei is 22.3\% (free neutron, proton and alpha particles have 24.3\%. 6.3\% and 47.1\% respectively).  Here the distribution of heavy nuclei is centered around a lower Z region and has multiple peaks in $n(Z)$ as plotted in Fig.~\ref{fig:z(n)}.  It is also interesting to note that in this case there is odd-even effect in the value of $Z$ for the abundances $n(Z)$.

It is instructive to compare the composition of matter in the Virial EOS with the existing EOS tables of Lattimer-Swesty and H Shen \etal. The location of the neutrinosphere in a supernova is sensitive to the composition of matter and is important for the emitted neutrino spectra. Studies of collective flavor oscillations of neutrinos during their streaming outside of the neutrinosphere have already indicated a sensitivity of neutrino oscillations to the emitted neutrino spectra from the neutrinosphere \cite{Duan10}. Below we will compare some examples for the composition of matter around the neutrinosphere from the Virial EOS, the L-S EOS, and S-S EOS.

Fig.~\ref{fig:compareabun} shows the mass fractions of neutrons, protons, alpha particles, and nuclei in matter at densities from 10$^{-6}$ to 10$^{-2}$ fm$^{-3}$. The matter has a temperature of 3.16 MeV and a proton fraction of 0.1 (a) or 0.3 (b), respectively. In panel (a) for a proton fraction of 0.1, free neutrons and protons dominate until the density reaches 10$^{-4}$ fm$^{-3}$ in all three EoSs.  Free nucleons dominate at low densities because of the high entropy.  Above 10$^{-4}$ fm$^{-3}$, alpha particles appear. The S-S EOS is close to our Virial results at densities below roughly 10$^{-3}$ fm$^{-3}$.   The L-S EOS significantly underestimates X$_\alpha$ and this may be due to an error in the alpha particle binding energy.   Alpha particles have larger abundances and exist up to higher densities in our Virial EOS than in the other EOSs.  This is partly because the attractive interactions between neutrons and alpha particles in the Virial expansion favors more alpha particles \cite{HS05}.  Heavy nuclei begin to appear around 4$\times$10$^{-4}$ fm$^{-3}$ in the L-S EOS, and at higher densities in the S-S EOS and our Virial EOS.  Moreover, the L-S EOS predicts the largest abundance for heavy nuclei, while ours predicts the smallest abundance.  Free neutrons have the largest abundance in our Virial EOS.  This is due to the strong attractive interaction between neutrons in the Virial expansion which lowers the energy and enhances the abundance of neutrons.  Note in the $Y_P$ = 0.1 case the Virial-Hartree transition happens at 0.0158 fm$^{-3}$. The right panel of Fig.~\ref{fig:compareabun}, for a different proton fraction of 0.3, has similar characteristics.  However here, alpha particles and heavy nuclei have much larger abundances than for the $Y_P$ = 0.1 case, since a higher proton fraction favors formation of nuclei. In this $Y_P$ = 0.3 case, the transition density from Virial gas to Hartree mean field calculations occurs at 6.3$\times$ 10$^{-3}$ fm$^{-3}$ as indicated by the dotted line in the figure.

In future work \cite{EOSIII} we will carefully interpolate the free energy results of this paper in a thermodynamically consistent way in order to accurately determine derivatives of the free energy with respect to temperature or density.  From these derivatives we will calculate a number of additional quantities such as the pressure or entropy.

\section{Summary and Outlook}
\label{summary}

In this paper we present our model for nuclear matter at
subnuclear density, the Virial expansion for a nonideal gas consisting
of neutrons, protons, alpha particles and thousands of heavy
nuclei.  We include second order virial corrections for light elements $A \leq$ 4, nuclear partition functions for heavy nuclei, and Coulomb corrections. At very
low density, the Virial expansion reduces to nuclear statistical
equilibrium. We calculate the free energy, and tabulate the
resulting EOS at over 73,000 grid points in the temperature
range $T$ = 0.158 to 15.8 MeV, the density range $n_B$ = 10$^{-8}$
to 0.1 fm$^{-3}$, and the proton fraction range $Y_P$ = 0.05
to 0.56. These calculations took over 1000 CPU days.
The above parameter space is complementary to that of the relativistic mean field results in our previous paper \cite{SHT10a}.

The treatment of Coulomb corrections in Wigner-Seitz approximation agrees reasonably with an analytical cluster expansion. Our results do not appear to be very sensitive to the mass table employed, or to the form of the partition function for heavy nuclei.  However, results at higher densities for mean field calculations are sensitive to the interaction employed.  In the future, we intend to match the Virial results of this paper to mean field calculations using a number of different interactions, see for example \cite{IUFSU}. 

Our EOS includes broad distributions of nuclei, that are not included in two commonly used EOS tables. These distributions of nuclei make the composition of nuclear matter in
our EOS different from L-S and S-S EOS tables.  These differences may be important for neutrino interactions. 

This paper provides the second part of our results for a complete EOS table that will cover a broad range of temperatures, densities, and proton fractions for use in supernova and neutron star merger simulations.  In the future we will use a thermodynamically consistent interpolation scheme to match the Virial EOS in this paper and the RMF EOS from our previous paper, or similar RMF models, and generate a complete EOS table \cite{EOSIII}.  Our Virial EOS is exact in the low density limit.  Finally the Virial expansion also provides a suitable framework for future work that includes other modern mass tables, such as HFB14 \cite{HFB14} or KTUY05 \cite{KTUY05}.

\section{Acknowledgement}

We thank Lorenz H$\ddot{\mathrm{u}}$edepohl, Thomas Janka, Jim Lattimer, Andreas Marek, Evan O'Connor, and Christian Ott for helpful discussions. This work was
supported in part by DOE grant DE-FG02-87ER40365. This material is based upon work supported by the National Science Foundation under Grants No. ACI-0338618l, OCI-0451237, OCI-0535258, OCI-0504075 and CNS-0521433.
This research was supported in part by Shared University Research grants from IBM, Inc. to Indiana University and by the Indiana METACyt Initiative.
The Indiana METACyt Initiative of Indiana University is supported in
part by Lilly Endowment.

\begin{widetext}

\begin{figure}[h]
 \centering
 \includegraphics[height=12cm]{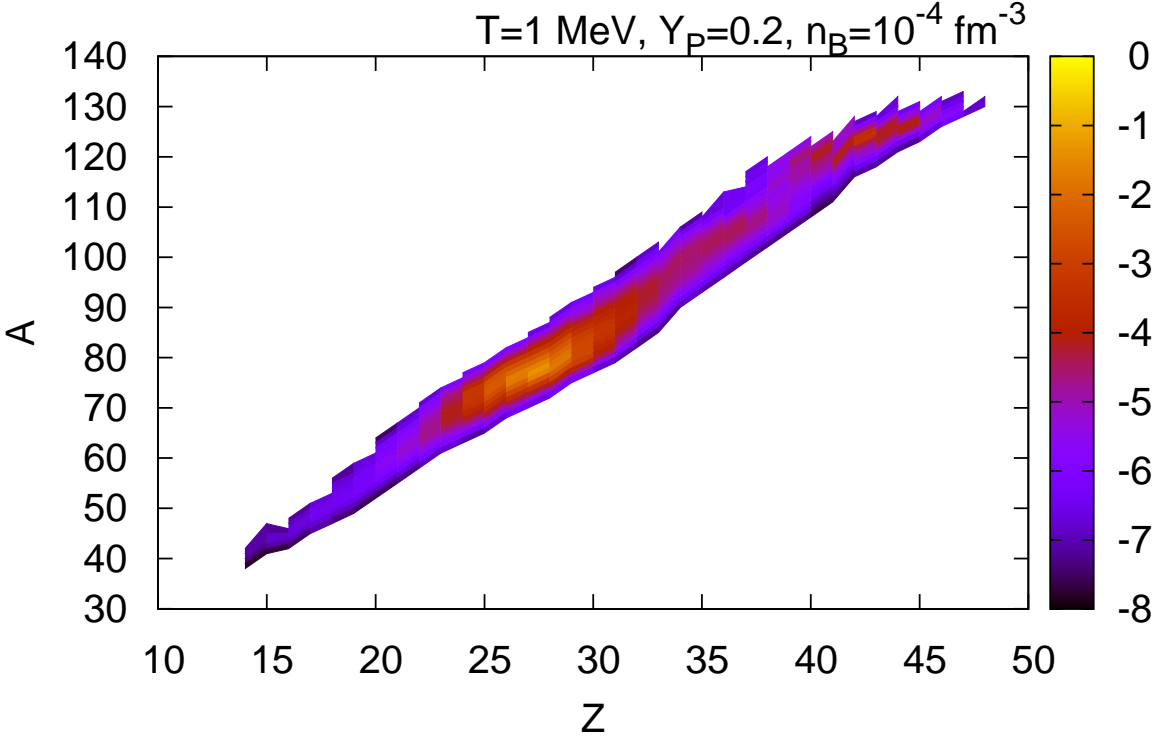}
 \includegraphics[height=12cm]{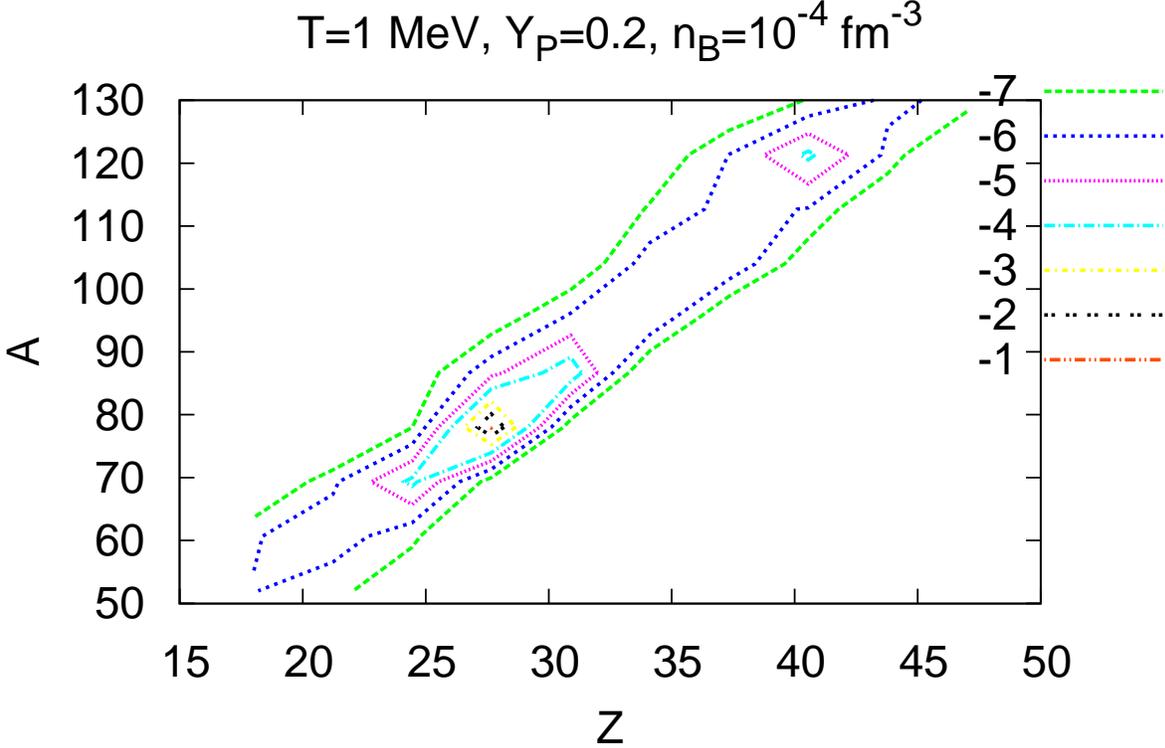}

\caption{(color on line) Mass fraction of nuclei in the nuclear chart for matter
at $T$ = 1 MeV, $n_B$ = $10^{-4}$
fm$^{-3}$, and $Y_P$ = 0.2. Upper panel: different colors indicate the mass fraction using a $ \rm{Log}_{10} $ scale. Lower panel: different lines indicate the contours of mass fraction in $ \rm{Log}_{10} $ scale from -1 to -7.}\label{fig:dist_T1Y2n-4}
\end{figure}

\begin{figure}[h]
 \centering
 \includegraphics[height=12cm]{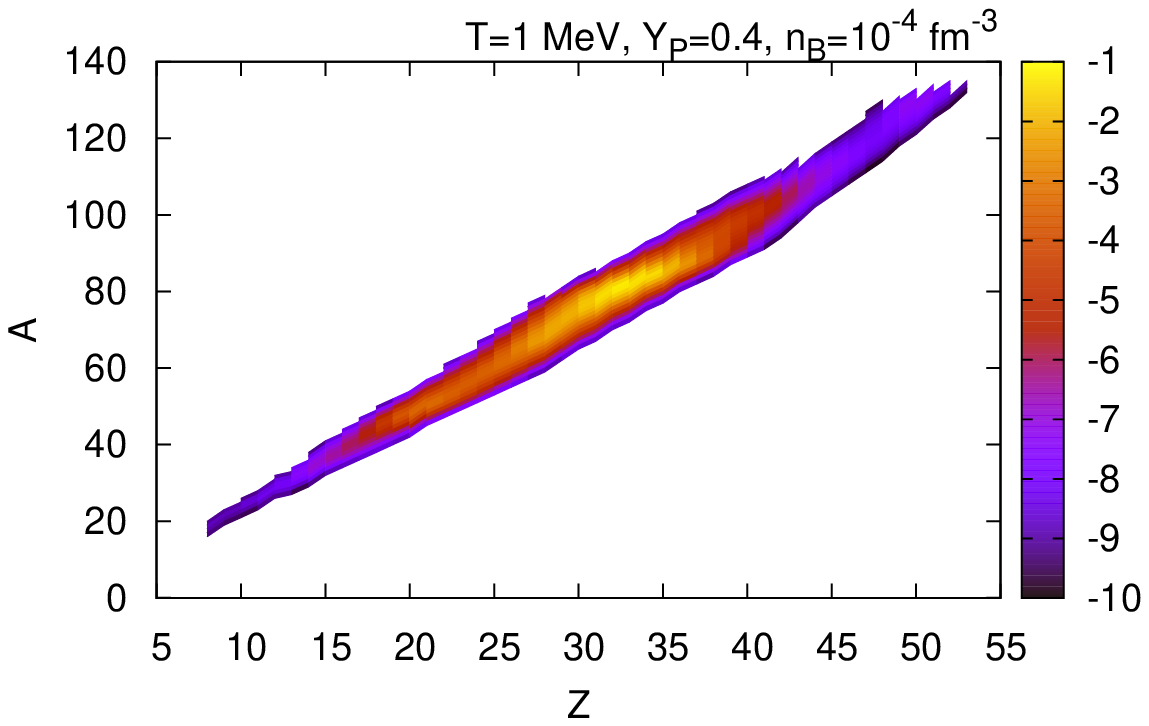}
 \includegraphics[height=12cm]{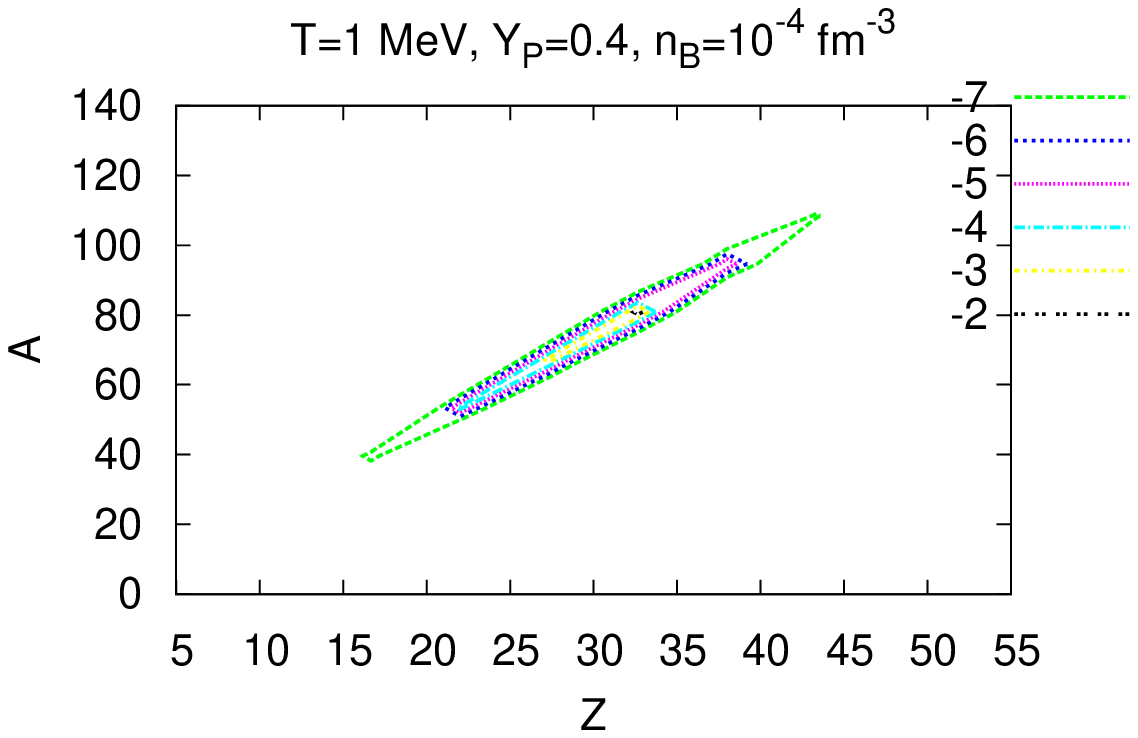}

\caption{(color on line) Mass fraction of nuclei in the nuclear chart for matter
at $T$ = 1 MeV, $n_B$ = $10^{-4}$
fm$^{-3}$, and $Y_P$ = 0.4. Upper panel: different colors indicate the mass fraction using a $ \rm{Log}_{10} $ scale. Lower panel: different lines indicate the contours of mass fraction in $ \rm{Log}_{10} $ scale from -2 to -7.}\label{fig:dist_T1Y4n-4}
\end{figure}

\begin{figure}[h]
 \centering
 \includegraphics[height=12cm]{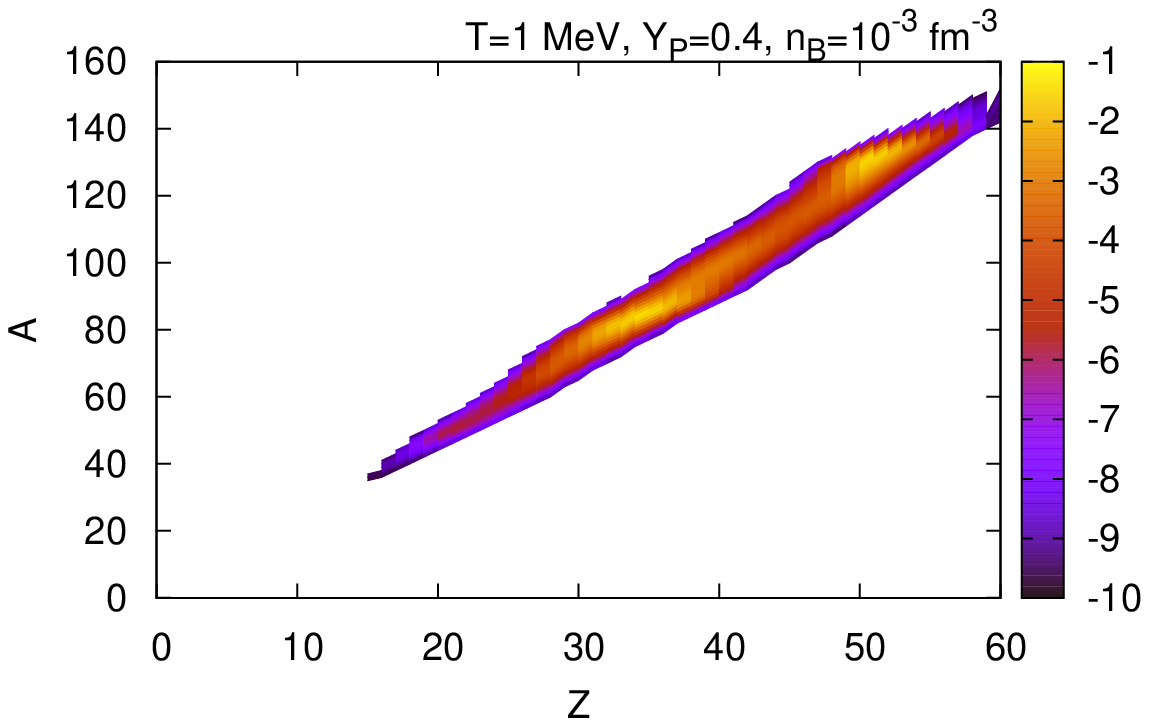}
 \includegraphics[height=12cm]{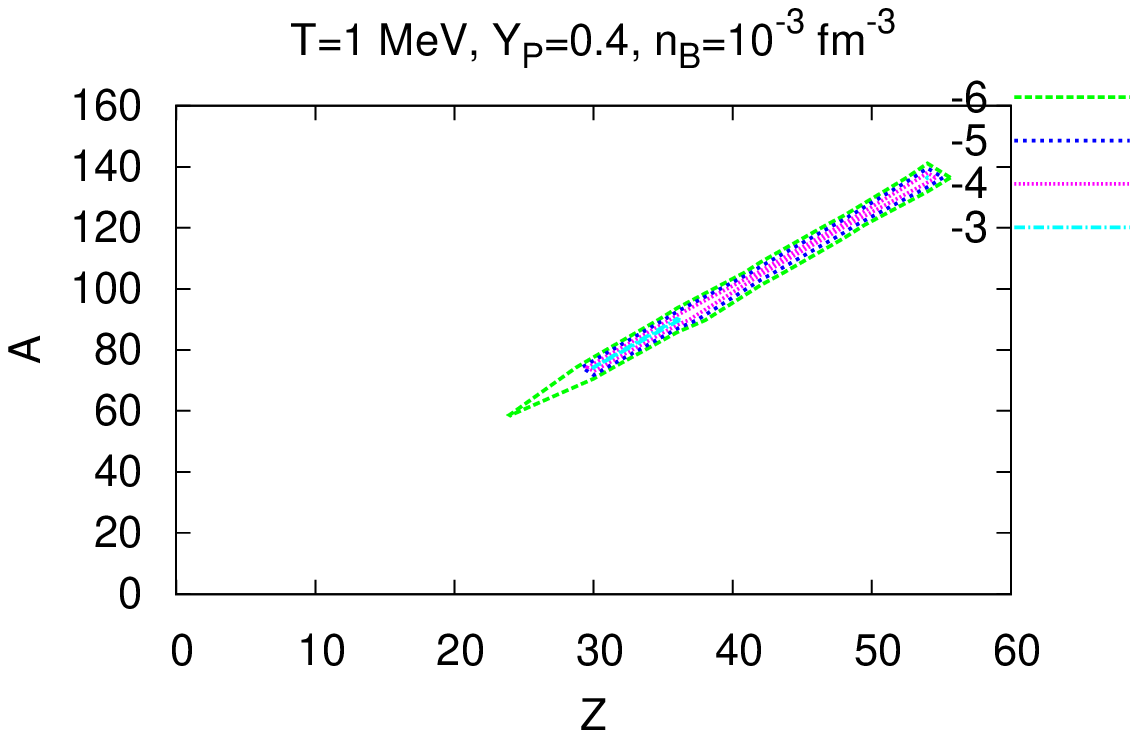}

\caption{(color on line) Mass fraction of nuclei in the nuclear chart for matter
at $T$ = 1 MeV, $n_B$ = $10^{-3}$
fm$^{-3}$, and $Y_P$ = 0.4. Upper panel: different colors indicate the mass fraction using a $ \rm{Log}_{10} $ scale. Lower panel: different lines indicate the contours of mass fraction in $ \rm{Log}_{10} $ scale from -3 to -6.}\label{fig:dist_T1Y4n-3}
\end{figure}

\begin{figure}[h]
 \centering
 \includegraphics[height=12cm]{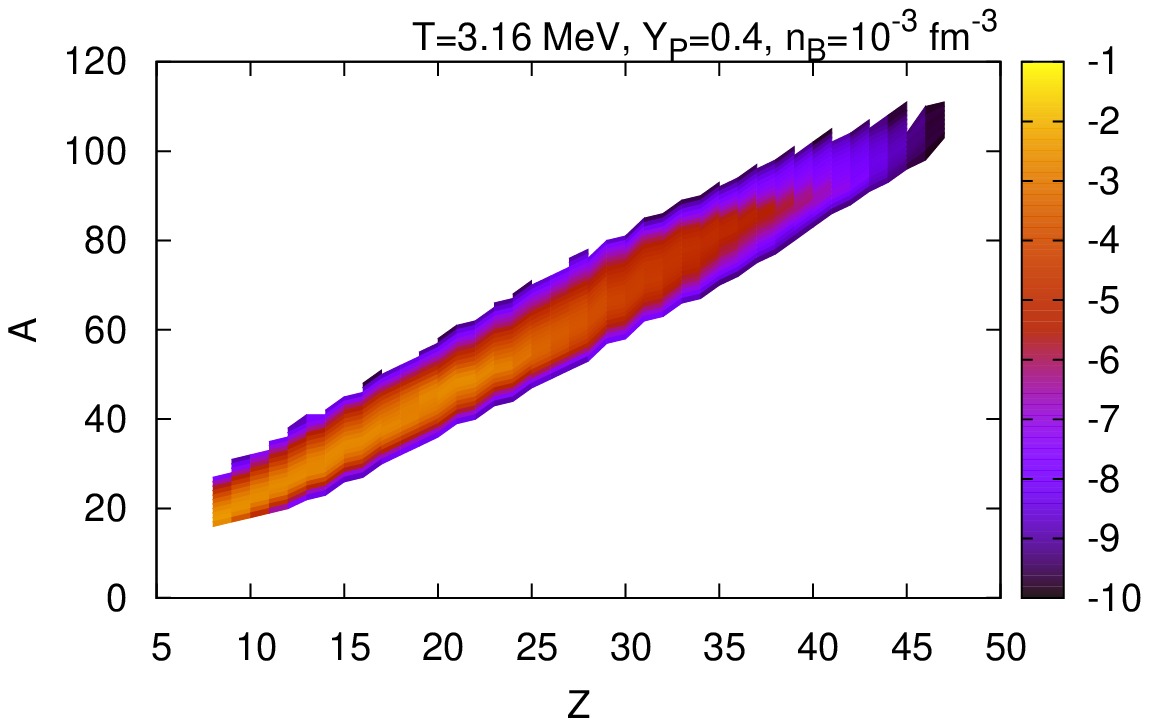}
 \includegraphics[height=12cm]{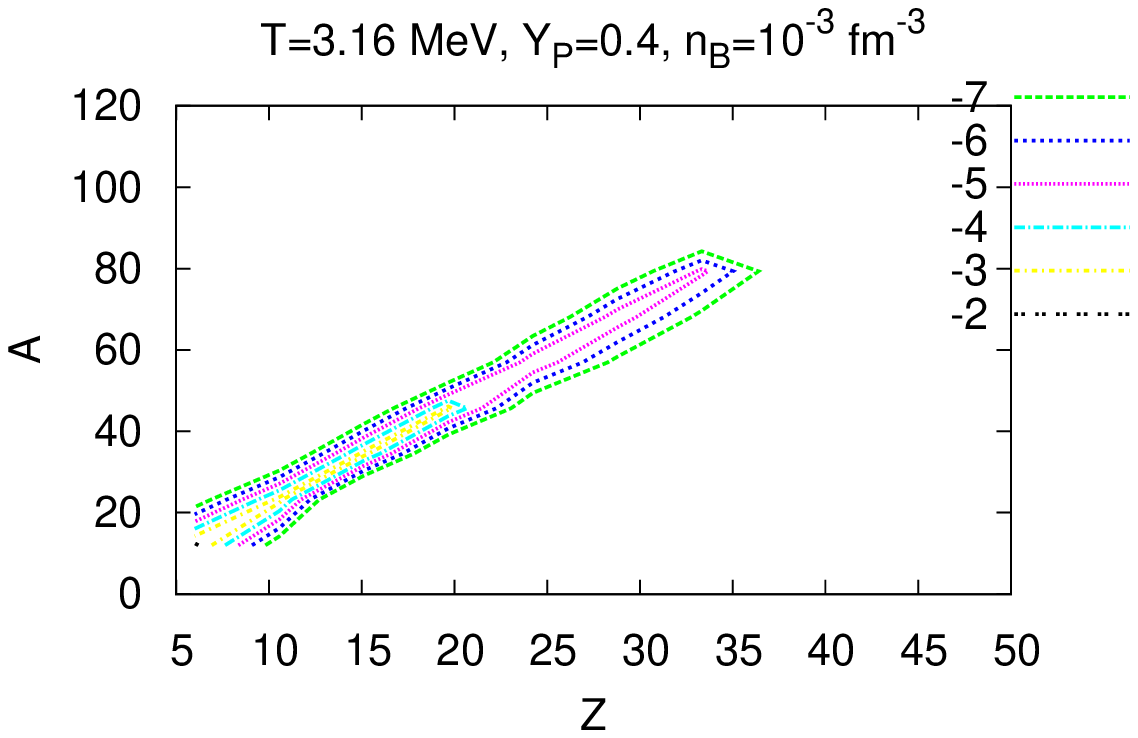}

\caption{(color on line) Mass fraction of nuclei in the nuclear chart for matter
at $T$ = 3.16 MeV, $n_B$ = $10^{-3}$
fm$^{-3}$, and $Y_P$ = 0.4. Upper panel: different colors indicate the mass fraction using a $ \rm{Log}_{10} $ scale. Lower panel: different lines indicate the contours of mass fraction in $ \rm{Log}_{10} $ scale from -2 to -7.}\label{fig:dist_T3Y4n-3}
\end{figure}


\begin{figure}[htbp]
 \centering
 \includegraphics[height=8.5cm,angle=-90]{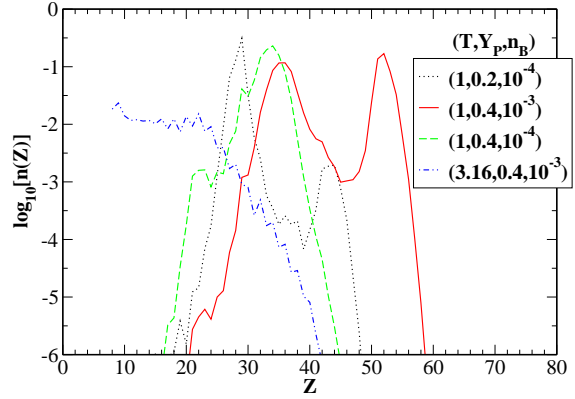}

\caption{(color on line) Mass fractions of nuclei $n(Z)$ for nuclei proton number $Z$, for the cases in Figs.~\ref{fig:dist_T1Y2n-4}, \ref{fig:dist_T1Y4n-4}, \ref{fig:dist_T1Y4n-3} and \ref{fig:dist_T3Y4n-3}. The triplet of values in the parenthesis are ($T$/[MeV], $Y_P$, $n_B$/[fm$^{-3}$]).}\label{fig:z(n)}
\end{figure}


\begin{figure}[h]
 \centering
 \includegraphics[height=8.5cm,angle=-90]{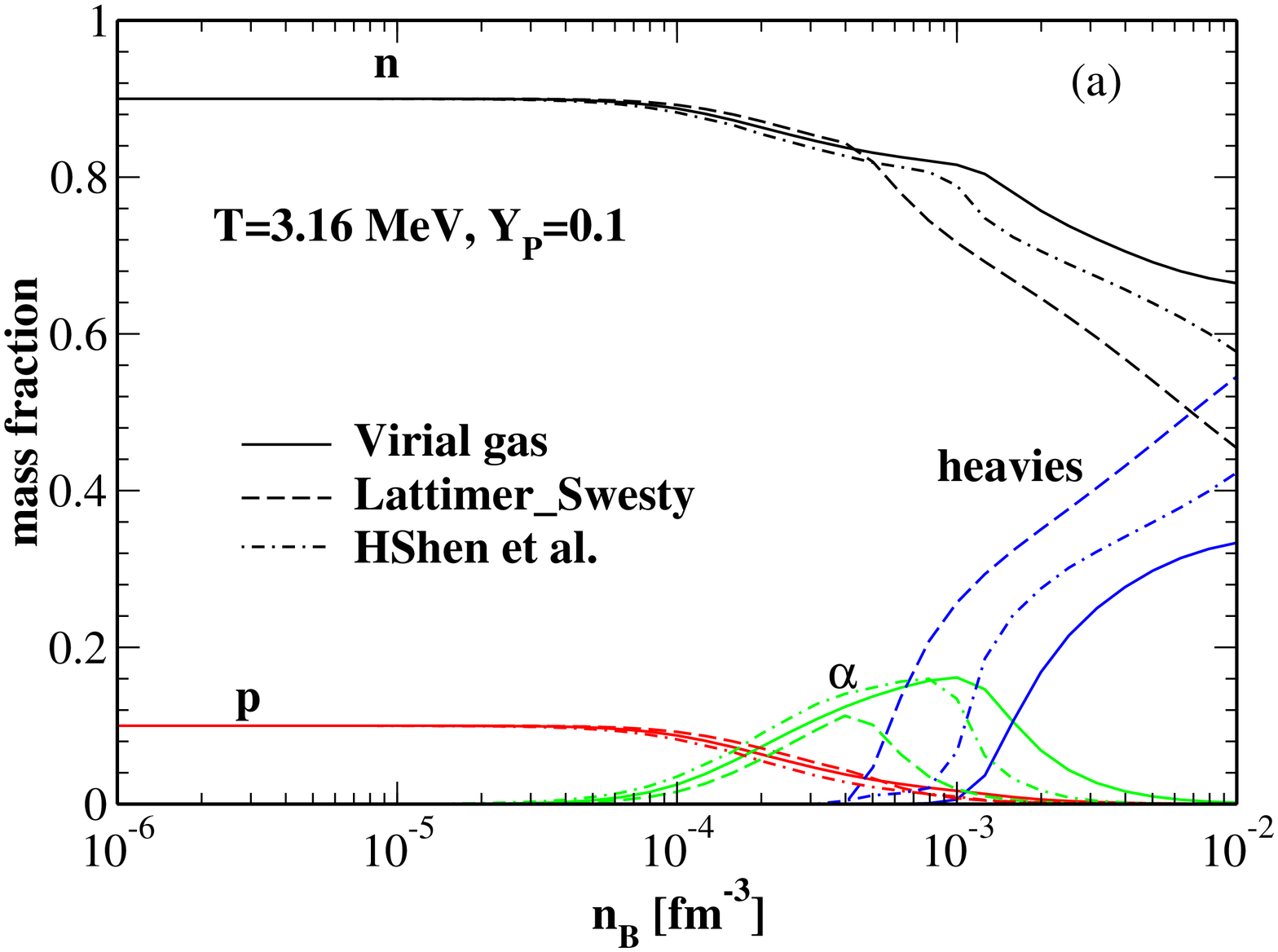}
 \includegraphics[height=8.5cm,angle=-90]{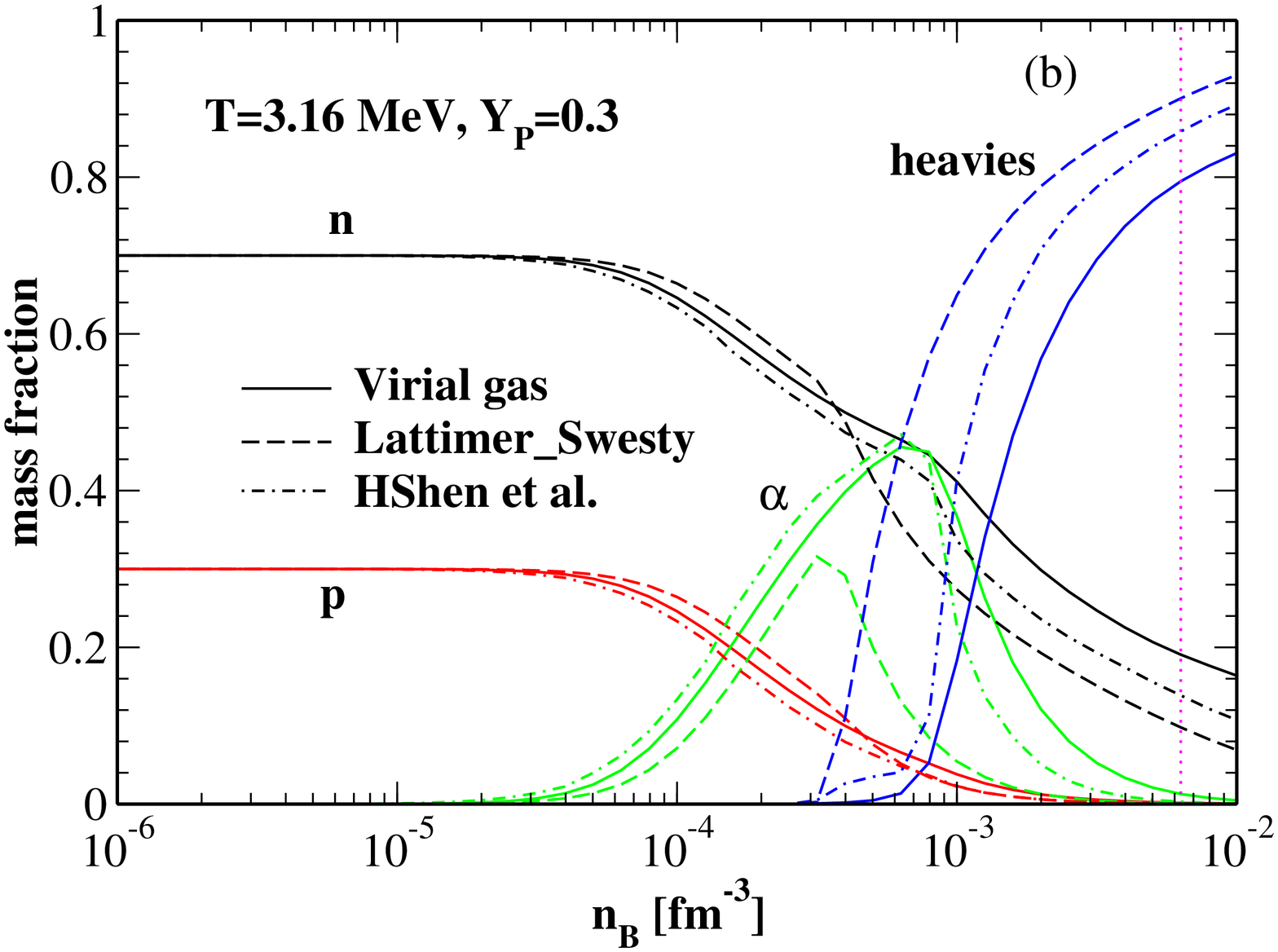}

\caption{(color on line) Mass fractions of matter
at $T$ = 3.16 MeV, $Y_P$ = 0.1 (a) and 0.3 (b).  The solid curves show our Virial EOS results, while the dashed lines show Lattimer-Swesty EOS results and the H Shen \etal\ EOS results are shown by the dot dashed curves. For $Y_P$ = 0.3, the dotted line indicates the transition density between Virial EOS and Hartree mean field results. For $Y_P$ = 0.1, the transition density between Virial EOS and Hartree is 0.0158 fm$^{-3}$.}\label{fig:compareabun}
\end{figure}
\end{widetext}


\begin{thebibliography}{99}


\bibitem{simulation1} M. Liebend\"{o}rfer, M. Rampp, H.-Th. Janka, and A. Mezzacappa, Astrophys. J. {\bf 620}, 840 (2005).

\bibitem{simulation2} R. Walder, A. Burrows, C. D. Ott, E. Livne,  I. Lichtenstadt, and M. Jarrah, Astrophys. J. {\bf 626}, 317 (2005).

\bibitem{Burrows04} A. Burrows, S. Reddy, and T. A. Thompson, Nucl. Phys. {\bf A 777}, 356 (2006).

\bibitem{SHT10a} G. Shen, C. J. Horowitz, and S. Teige, Phys. Rev. C {\bf 82}, 015806 (2010).

\bibitem{FRDM} P. M\"{o}ller, J. R. Nix, and W. J. Swiatecki, Atom.
Data Nucl. Data Tables, {\bf 59}, 185 (1995).

\bibitem{LS} J. M. Lattimer and F. D. Swesty, Nucl. Phys. {\bf A
535}, 331 (1991).

\bibitem{Shen98a} H. Shen, H. Toki, K. Oyamatsu, and K. Sumiyoshi,
Nucl. Phys. {\bf A 637}, 435 (1998).

\bibitem{Shen98} H. Shen, H. Toki, K. Oyamatsu, and K. Sumiyoshi,
Prog. Theo. Phys. {\bf 100}, 1013 (1998).


\bibitem{SN1987a} M. Costantinit, A. Ianni, and F. Visanni, Phys. Rev. D {\bf 70}, 043006 (2004);
C. Lunardini and A.Y. Smirnov, Astropart. Phys. {\bf 21}, 703 (2004).

\bibitem{light07} A. Arcones, G. Mart\'{i}nez-Pinedo, E. O'connor, A. Schwenk, H.-T. Janka, C. J. Horowitz, and K. Langanke, Phys. Rev. C
{\bf 78}, 015806 (2008).


\bibitem{NSE} B.S. Meyer, Ann. Rev. Astron. Astrophys. {\bf 32}, 153 (1994).

\bibitem{NSE09} S. I. Blinnikov, I. V. Panov, M. A. Rudzsky, and K.
Sumiyoshi, arXiv: 0904.3849.

\bibitem{Hempel09} Matthias Hempel and J\"{u}rgen Schaffner-Bielich,
arXiv:0911.4073v1.

\bibitem{unitarygas} See for example, J. Carlson, S.-Y. Chang, V. R. Pandharipande, and K. E. Schmidt,
Phys. Rev. Lett. {\bf 91}, 050401 (2003).

\bibitem{Ho04} T.-L. Ho and E. J. Mueller, Phys. Rev. Lett. {\bf
92}, 160404 (2004).

\bibitem{HS05b} C. J. Horowitz and A. Schwenk, Phys. Lett. {\bf B638}, 153 (2006).




\bibitem{HS05} C. J. Horowitz and A. Schwenk, Nucl. Phys. {\bf A 776}, 55 (2006).

\bibitem{mass3} C.~J.~Horowitz and A. Schwenk, in preparation.

\bibitem{Fowler78} W. A. Fowler, C. A. Engelbrecht, and S. E.
Woosley, ApJ {\bf 226}, 984 (1978).

\bibitem{Rauscher97} T. Rauscher, F.-K. Thielemann, and K.-L.
Kratz, Phys. Rev. {\bf C 56}, 1613 (1997).

\bibitem{Gilbert65} A. Gilbert and A. G. W. Cameron, Canadian Journal of Physics. {\bf
43}, 1446 (1965).

\bibitem{Mazurek79} T. J. Mazurek, J. M. Lattimer, and G. E.
Brown, Astrophysical Journal {\bf 229}, 713 (1979).

\bibitem{Engelbrecht91} C. A. Engelbrecht and J. R. Engelbrecht,
Ann. Phys. {\bf 207}, 1 (1991).

\bibitem{Bethe36} H. A. Bethe, Phys. Rev. {\bf 50}, 332 (1936).

\bibitem{Rauscher00} T. Rauscher and F.-K. Thielemann, Atom. Data Nucl. Data Tab. {\bf 75}, 1
(2000).

\bibitem{mc} D. J. Dean, S. E. Koonin, K.-H. Langanke, P. B. Radha, and Y.
Alhassid, Phys. Rev. Lett. {\bf 74}, 2909 (1995).

\bibitem{Paar97} V. Paar and R. Pezer, Phys. Rev. C {\bf 55}, R1637 (1997).

\bibitem{coulomb} N. V. Brilliantov, Contrib. Plasma Phys. {\bf 38}, N4
(1998).

\bibitem{BBP} G. Baym, H. A. Bethe, and C. J. Pethick, Nucl. Phys.
{\bf A 175}, 225 (1971).

\bibitem{Nadyozhin04} D. K. Nzdyozhin and A. V. Yudin, Astro.
Lett. {\bf 30}, 634 (2004).

\bibitem{HFB14} S. Goriely, M. Samyn, and J. M. Pearson, Phys. Rev. C {\bf 75},
064312 (2007).

\bibitem{Duan10} See for a recent review, H. Duan, G. M. Fuller, and Y.-Z. Qian, arXiv:1001.2799.

\bibitem{EOSIII}G. Shen, C. J. Horowitz, and S. Teige, to be published.

\bibitem{IUFSU}     F. J. Fattoyev, C. J. Horowitz, J. Piekarewicz, and G. Shen, arXiv:1008.3030.

\bibitem{KTUY05} H. Koura, T. Tachibana, M. Uno, and
M. Yamada, Prog. Theo. Phys. {\bf 113}, 305 (2005).

\end{thebibliography}
\end{document}